\documentclass[a4paper,11pt]{article}
\usepackage{geometry}
\usepackage[utf8]{inputenc}
\usepackage{amsmath}
\usepackage{amssymb}
\usepackage[
    backend=biber,
    style=numeric,
    sorting=none,
    url=false,
]{biblatex}
\usepackage{doi}

\usepackage{tikz}
\usepackage{pgfplots}
\usepackage{nicefrac}
\usepackage[ruled]{algorithm2e}
\usepackage{appendix}
\usepackage{longtable}
\usepackage{url}
\usepackage{graphicx}
\usepackage{hyperref}

\usetikzlibrary{positioning, calc, shapes.misc, external}
\usepgfplotslibrary{external}
\pgfplotsset{compat=1.18}
\usepackage{nicefrac}
\usepackage{xcolor}
\usepackage{bm}

\usepackage{caption}
\usepackage{subcaption}
\usepackage[normalem]{ulem}

\usepackage{tabularray}

\usepackage{authblk}

\renewcommand{\vec}[1]{\bm{#1}}
\newcommand{\ten}[1]{\bm{#1}}

\title{A hyperreduced manifold learning approach to nonlinear model order reduction for the homogenisation of hyperelastic RVEs}

\author{Erik Faust\thanks{erik.faust@mv.rptu.de} }
\author{Lisa Scheunemann\thanks{lisa.scheunemann@mv.rptu.de}}
\affil[]{Chair of Applied Mechanics, Department of Mechanical and Process Engineering, RPTU Kaiserslautern-Landau}

\begin{document}

\maketitle

\begin{abstract}
In a recent work, we proposed a graph-based manifold learning scheme for the nonlinear Galerkin-reduction of quasi-static solid mechanical problems~\cite{SchFau:2024:mla}. The resulting nonlinear approximation spaces can closely and flexibly represent nonlinear solution manifolds. 
{The present work discusses how this nonlinear model order reduction (MOR) approach can be employed to reduce online computational costs by multiple orders of magnitude while retaining high levels of accuracy. We integrate two popular hyperreduction methods into the nonlinear MOR framework and discuss how we achieve an algorithmic complexity which is independent from the original system size.}
Furthermore, improvements are made to the local online linearisation scheme for the sake of performance and robustness. 
On an example RVE problem, the MOR scheme accelerates computations by more than two orders of magnitude with little training data and negligible loss of accuracy. Additionally, the algorithm Pareto-dominates alternative approaches in the trade-off between accuracy and runtime on the considered example.
\end{abstract}

\section{Introduction}

Engineering tasks ranging from uncertainty quantification~\cite{BenGugWil:2015:spm} over design optimisation~\cite{GanZab:2004:dls,ChaAntBuf:2023:lrb,ChaAntBuf:2024:fpa}, the tackling of inverse problems~\cite{Bha:2017:rom,BolBul:2011:ect,GhaWil:2021:lpm,BulMai:2011:pod,GarMaiNov:2012:coe} and simple parameter studies to multiscale modeling~\cite{FriBoh:2013:rbh,DenSodApe:2022:rmm,GuoRokVer:2024:rom,BhaMat:2016:nmr} necessitate repeated evaluations of parameterised simulation models. When an engineering task requires hundreds or even thousands of queries to such a simulation model, the use of traditional high-fidelity simulation techniques, e.g. the Finite Element Method (FEM), may result in prohibitive computational costs. The parametric setting, however, permits the use of parametric 
model order reduction (MOR) techniques: a reduced order model (ROM) can be constructed based on empirical \textit{snapshot} data and a priori knowledge about quantities of interest in an offline training phase. The ROM can then be deployed at reduced computational cost in the online application phase.

In this article, we primarily deal with the acceleration of computations on representative volume elements (RVEs) in the context of quasi-static, hyperelastic multiscale modeling, though the methods deployed to this end are {intended to be} sufficiently general to be applied in different parametric settings as well. In a previous work~\cite{SchFau:2024:mla}, we outlined a range of approaches to accelerating solid-mechanical multiscale simulations: these include simplified physical models~\cite{SchBalBra:2015:d3s,DycHub:2023:dmm}, methods which simplify system equations by exploiting the specific physics~\cite{MicSuq:2003:ntf,FriLeu:2013:rbh,CovDeFri:2018:cro,LiuBesLiu:2016:sca,ElGonChi:2013:fml,NirAlfGon:2013:mor}, projection-based approaches which simplify equations in a general, data-based manner~\cite{HerOliHue:2014:hmr,ZahAveFar:2017:mpb,HerCaiFer:2017:dhn,SolBraZab:2017:nsd,RocKerVan:2020:msm,RasLloHue:2021:hpr,LanHutKie:2024:mhr,GuoRokVer:2024:rom,WulHau:2025:ech}, and data-based surrogate models~\cite{BhaMat:2016:nmr,BhaMat:2020:ndr,KimShi:2024:dmf,RocKerVan:2020:msm}. Which approach is most appropriate strongly depends on the problem at hand. The choice of method is generally informed by trade-offs between that method's offline training cost and data hungriness, online accuracy, online computational cost, robustness and generality, as well as the ease of use and implementation.

As discussed in~\cite{SchFau:2024:mla}, projection-based MOR techniques strike an attractive balance between these performance characteristics: they require very little data and enable considerable reductions in computational cost while maintaining high degrees of accuracy, robustness and generality. This is achieved at the cost of intrusiveness and comparatively high implementational complexity\footnote{Note that, in this work, we take ``projection-based MOR" to only denote intrusive methods which to some extent retain the integration of the local, underlying balance laws and thus remain closely tied to the underlying physics. For non-intrusive techniques which strike a different balance between the aforementioned performance criteria, see e.g.~\cite{GeeWriWil:2023:oin,CicFreGuo:2023:uqn,CicFreMan:2022:ddl,FleDeAta:2025:nph}.}.
{Important structural features of the underlying full-order simulation scheme are retained, while the algorithmic complexity of expensive operations, such as large linear system solves and integration of balance laws over finely resolved calculation domains, is significantly reduced.}

Conceptually, projection-based MOR capitalises on the observation that solutions $\vec{u}(\vec{p})\in \mathbb{R}^D$ to a well-posed, quasi-static, parametric problem $\vec{g}(\vec{u};\vec{p})=\vec{0}$ which is discretised in terms of $D$ unknowns and formulated in terms of parameters $\vec{p}\in \mathbb{R}^\delta$ lie on a $\delta \ll D$-dimensional solution manifold $\mathcal{M}_{\vec{u}}$.
Projection-based ROMs attempt to look for solutions not in the high-dimensional solution space $\mathbb{R}^D$, but a $d\ll D$-dimensional approximation space (or approximation manifold) $\mathcal{M}_{\vec{\bar{u}}}$, which in the ideal case is a tight superset of the low-dimensional, but possibly strongly nonlinear solution manifold. The Galerkin projection of the unknown variables and system equations onto this approximation space can significantly reduce the computational cost of solving the linear equation systems arising from, e.g. a Newton-Raphson solution scheme.

Intrusive projection-based MOR techniques additionally rely on a hyperreduction methodology to reduce the computational cost of integrating the underlying local balance laws to assemble the aforementioned linear equation systems~\cite{Ryc:2009:hrm}. These methods generally restrict the evaluation of the underlying local physical equations to a restricted set of mesh entities such as residual degrees of freedom~\cite{EveSir:1995:klp,Wil:2006:ufls,NguPatPer:2008:bpi,AstWeiWil:2008:mpe,ChaSor:2010:nmr,TisDedRix:2013:dmd,LauCheShi:2024:sps,Ryc:2009:hrm,CarBouFar:2011:enl,CarFarCor:2013:gmn,CarBarAnt:2017:glp,CheJiNar:2021:lrc}, elements~\cite{FarAveCha:2014:drn}, Gauss points~\cite{HerCaiFer:2017:dhn} or, more abstractly, points in strain space~\cite{Wul:2024:sch,WulHau:2025:ech}. The contribution of mesh entities which are discarded in favour of computational speed can then be approximated explicitly or implicitly. This low-order approximation is possible because the reduced residual is a function of the location of the sought solution on the solution manifold $\vec{u} \in \mathcal{M}_{\vec{u}}$, as well as of the location of the current iterate in the approximation space $\vec{\bar{u}} \in \mathcal{M}_{\vec{\bar{u}}}$ -- and thus also lies on a low-dimensional manifold.

{Hyperreduced, projection-based MOR algorithms can achieve truly significant reductions in runtime and memory costs when care is taken that their computational complexity is no longer dependent on the size of the underlying full-order problem. In the following, we outline established Galerkin- and hyperreduction techniques which can be utilised to this end. Additionally, further discussion of algorithmic complexity is integrated into the remainder of this work.
}

In the context of intrusive, projection-based MOR for quasi-static solid mechanics, the Proper Orthogonal Decomposition (POD) is a popular approach to performing the Galerkin projection. 
It defines the approximation space as a linear subspace of the original solution space~\cite{YvoHe:2007:rmm,MonYvoHe:2008:chn,RadBedSti:2016:dmm,Wei:2019:tpo}. While the POD is attractive on account of its simplicity and robustness, relatively high model sizes $d>\delta$ might be required to accurately represent a strongly nonlinear solution manifold. 
The local basis method (LPOD) thus aims for a tighter representation of the solution manifold by localising the POD to regions of solution- or parameter space~\cite{AmsZahFar:2012:nmo,AmsZahWas:2015:flr,DanCasAkk:2020:mor,DanCasAkk:2022:pca,DanCasAkk:2022:uqi,ChaAntBuf:2023:lrb,ChaAntBuf:2024:fpa}. Alternatively, polynomial manifold (PM) approaches attempt to achieve a smoother, tighter approximation via a polynomial Ansatz~\cite{JaiTisRut:2017:qmm,JaiTis:2019:hnm,GeeWriWil:2023:oin,BarFar:2022:qam}. Artificial Neural network (ANN) approximation spaces (using Feedforward Neural Networks or Autoencoders) have also been used with a view to improved flexibility and generality, but are relatively data-hungry~\cite{KimChoWid:2022:fap,FreDedMan:2021:cdl,FreMan:2022:ped,BarFarMad:2023:npma}. 

In a recent proof-of-concept~\cite{SchFau:2024:mla}, we instead proposed a graph-based manifold learning (GML) approach for the nonlinear Galerkin-reduction of quasi-static solid-mechanical problems.
GML techniques such as Locally Linear Embedding (LLE)~\cite{RowSau:2000:ndr} strike an attractive balance between the flexibility of an ANN Ansatz and the data-economy of the POD. As a result, they have successfully been applied to the surrogate modeling of RVEs~\cite{BhaMat:2016:nmr,BhaMat:2020:ndr,KimShi:2024:dmf} and fluid problems~\cite{Men:2023:lnd} as well as to non-intrusive model reduction in elastodynamics~\cite{MilArr:2013:nml}. As opposed to the approaches outlined above, the LLE and related methods do not utilise an explicit Ansatz for the approximation space, instead only providing a low-dimensional embedding of the snapshot data. Approximation spaces have to be constructed a posteriori. \cite{MilArr:2013:nml} used max-Ent interpolants to this end, \cite{DieMuiZlo:2021:ndr} a LPOD-like linearisation, and~\cite{SchFau:2024:mla} a forward-Euler-like scheme.

While research on nonlinear approximation spaces is quite well-developed for the dynamic case, particularly in fluid mechanics~\cite{PenMoh:2014:nmr,AmsZahFar:2012:nmo,AmsZahWas:2015:flr,DanCasAkk:2022:uqi,SimVlaGar:2023:vnv,JaiTis:2018:shg,SchGugPeh:2024:esr,JaiTis:2019:hnm,BarFar:2022:qam,KimChoWid:2022:fap,CocTenRiz:2023:hae,RomStaRoz:2025:eho,LeeCar:2020:mrd,RomStaRoz:2022:nmr,RomStaRoz:2023:eho,BarFarMad:2023:npma,DieMuiZlo:2021:ndr}, their application to (quasi-static) solid mechanics in general and computational homogenisation in particular are in a stage of relative infancy.
Dynamic solid problems have been tackled with LPOD approaches in~\cite{PagManQua:2018:nap,DanCasAkk:2022:uqi,VlaTatAga:2021:lba,SimVlaGar:2023:vnv}, with the PM in~\cite{JaiTis:2018:shg,JaiTis:2019:hnm}, and with a (non-intrusive) manifold learning scheme in~\cite{MilArr:2013:nml}. The LPOD has also been applied to quasi-static solid mechanics in~\cite{GhaTisSim:2017:pdm,ChaAntBuf:2024:fpa} and a time-dependent computational homogenisation problem in~\cite{HeAveFar:2020:sar}. The PM has been applied to a quasi-static problem by~\cite{ZhaZhaRee:2024:nmm}, but its performance in high-dimensional parameter spaces and in combination with hyperreduction techniques remain to be explored. 
Similarly, we applied a proof-of-concept GML approach to an RVE problem in~\cite{SchFau:2024:mla}, but the hyperreduction, as well as the homogenisation of this approach remain to be addressed.

It is important to note that the context of employment results in some specific challenges for nonlinear MOR techniques: in quasi-static applications, the search for a static solution, rather than the evolution of a dynamic one, is constrained to the approximation space~\cite{SchFau:2024:mla}. Data can be scarce, and the accurate approximation of the reduced system Jacobian, as well as the convergence of the reduced scheme, are critical. In the context of multiscale modeling, the parameter space can relatively large (at a minimum, $\delta=6$), and an efficient and accurate post-processing of homogenised stresses is paramount.

By comparison, there has been extensive research on hyperreduction techniques in general and for the quasi-static solid case in particular. Approaches which assemble a reduced set of entries of the residual vector and approximate the reduced residual in an approximated Galerkin scheme include the Gappy POD~\cite{EveSir:1995:klp,Wil:2006:ufls}, Best Points Interpolation~\cite{NguPatPer:2008:bpi}, Missing Point Estimation~\cite{AstWeiWil:2008:mpe}, Discrete Empirical Interpolation Method (DEIM)~\cite{ChaSor:2010:nmr}, unassembled DEIM~\cite{TisDedRix:2013:dmd}, and S-OPT~\cite{LauCheShi:2024:sps}. While these approaches are attractive on account of their simplicity, a lack of preserved structure in the resulting reduced equation systems has been noted to result in suboptimal robustness~\cite{SolBraZab:2017:nsd,BraDavMer:2019:rmh}. Methods for structure-preservation have been proposed for specific problem classes (usually in dynamics)~\cite{CarTumBog:2015:pls,ChaBeaGug:2016:smr,PagVis:2023:ghn}, but this may come at the cost of reduced generality and increased implementational effort.
In contrast, collocation-like Petrov-Galerkin methods solve the underlying equation systems at the selected entries of the residual (either in a weighted or least-squares manner) more directly. Such methods include the eponymous hyperreduction (HR) technique from~\cite{Ryc:2009:hrm}, Gauss-Newton with Approximate Tensors (GNAT)~\cite{CarBouFar:2011:enl,CarFarCor:2013:gmn}, Least-Squares Petrov Galerkin (LSPG)~\cite{CarBarAnt:2017:glp}, and Reduced Over-Collocation~\cite{CheJiNar:2021:lrc}. Some of these methods strike a promising balance between simplicity, robustness~\cite{SolBraZab:2017:nsd,BraDavMer:2019:rmh} and generality.
In the context of the FEM, Energy Preserving Weighting and Sampling (ECSW), which considers the contribution of a restricted set of elements to the reduced residual, provides a natural, robust, structure-preserving alternative~\cite{FarAveCha:2014:drn}. The Empirical Cubature Method generalises this approach to integration points and augments ECSW with additional physical constraints~\cite{HerCaiFer:2017:dhn}. 
Finally, promising strain-based hyperreduction techniques such as the Empirically Corrected Cluster Cubature were recently proposed in~\cite{Wul:2024:sch,WulHau:2025:ech,WulHau:2025:ech}. These approaches select integration points at which to evaluate the material law as statistical representatives of the set of integration points in strain space and allow for natural and highly efficient hyperreduction especially in the context of computational homogenisation.

As mentioned above, such hyperreduction techniques are usually applied in tandem with the POD in the context of computational homogenisation. For example, approaches assembling a reduced set of residual degrees of freedom are applied in~\cite{GouKerBor:2014:bac,HerOliHue:2014:hmr,GhaYanGil:2015:fmr,OhlRavSch:2016:mrm,SolBraZab:2017:nsd,VanRemGee:2018:iem,BraDavMer:2019:rmh,SoLee:2025:tsh} while a reduced set of elements are assembled by~\cite{ZahAveFar:2017:mpb,HerCaiFer:2017:dhn,VanRemGee:2018:iem,CaiMroTor:2019:hpra,RocKerVan:2020:msm,RasLloHue:2021:hpr,LanHutKie:2024:mhr,GuoRokVer:2024:rom,LanHutKie:2025:ehe,GuoKouGee:2025:rms}, and~\cite{KunFri:2019:fsh,Wul:2024:sch,WulHau:2025:ech,WulHau:2025:ech} apply strain-based approaches. Additionally, the LPOD has been coupled with an element-based approach by~\cite{HeAveFar:2020:sar}. In quasi-static solid mechanics more generally, the POD has been applied with residual-based hyperreduction approaches in~\cite{RadRee:2016:pbm,GhaTisSim:2017:pdm,FriHaaRyc:2018:ach,BraDavMer:2019:rmh,KanWeiHe:2021:hcm,CicFrePag:2022:pro,KasKehBre:2023:dei,AgoArgBer:2024:prma,BonManQua:2017:mdt} and element approaches in~\cite{GouDur:2018:fgr,PhaBreBou:2020:rmf,IolSamTad:2021:pmr,LeeLeeCho:2021:rmn,TraMarMai:2024:ehh}. Furthermore, the LPOD has been deployed on quasi-static problems with DEIM-like approaches by~\cite{GhaTisSim:2017:pdm,PagManQua:2018:nap,ChaAntBuf:2024:fpa}. On dynamic problems, the LPOD has also been variously applied with the DEIM~\cite{DanCasAkk:2022:uqi}, gappy POD~\cite{DanCasAkk:2022:uqi}, GNAT~\cite{AmsZahFar:2012:nmo,AmsZahWas:2015:flr}, and ECSW~\cite{VlaTatAga:2021:lba,SimVlaGar:2023:vnv}. The PM, meanwhile, has been coupled with DEIM-like approaches~\cite{JaiTis:2018:shg,SchGugPeh:2024:esr}, and the ECSW~\cite{JaiTis:2019:hnm,BarFar:2022:qam}. Meanwhile, ANN approaches have been deployed with
S-OPT~\cite{RomStaRoz:2025:eho}, HR~\cite{CocTenRiz:2023:hae}, GNAT~\cite{KimChoWid:2022:fap,RomStaRoz:2025:eho}, LSPG~\cite{LeeCar:2020:mrd,RomStaRoz:2022:nmr,RomStaRoz:2023:eho}, and ECSW~\cite{BarFarMad:2023:npma,RomStaRoz:2025:eho}. 
There have also been some efforts to exploit nonlinear correlations in the hyperreduction step itself, see e.g.~\cite{OttRow:2019:dei,MarWev:2019:phn,HirPicHes:2024:nei}.

In this work, we build on the proof-of-concept GML approach to nonlinear MOR proposed in~\cite{SchFau:2024:mla}. 
{In order to reduce the computational cost of assembling reduced equation systems, we augment the scheme with two popular hyperreduction techniques -- namely, the DEIM and the LSPG. We emphasise aspects of the implementation which are necessary to achieve significant reductions in runtime and memory costs, and discuss algorithmic complexity where appropriate.}
Additionally, we extend the nonlinear MOR scheme to allow for the efficient computation of homogenised stresses and stiffnesses for multiscale applications.
Finally, we compare the performance of the resulting hyperreduced nonlinear MOR scheme on a simple example RVE problem with several alternative techniques: the POD, LPOD, and PM. 
On the example RVE problem, the hyperreduced GML approach is shown to yield speedups of two orders of magnitude with respect to full FEM simulations while retaining high levels of accuracy. Furthermore, an LLE approach is shown to be able to Pareto-dominate alternative methods in terms of the tradeoff between computational efficiency, accuracy, and robustness.

The remainder of this work is structured as follows: Sections~\ref{s:mech} and~\ref{s:RVE} outline the fundamentals of the underlying quasi-static solid mechanical- and computational homogenisation problems, respectively. Section~\ref{s:NLMOR} then discusses the nonlinear projection-based MOR techniques to be investigated, including their application to the Galerkin-reduction of a Newton-Raphson solver scheme and to the homogenisation of stress and stiffness. Section~\ref{s:HR} covers the hyperreduction techniques. Finally, Section~\ref{s:results} features numerical comparisons of the MOR and hyperreduction techniques on an example RVE problem.

\section{Quasi-static solid-mechanical problems}\label{s:mech}

On a solid domain $\Omega$ consisting of points $\vec{X}$ in the reference configuration, the weak form of the quasi-static balance of linear momentum can be written as
\begin{equation}
    \int_{\Omega} \ten{P} : \delta {\ten{F}} dV  - \int_{\Omega} \vec{b} \cdot \delta \text{\textbf{u}} dV - \int_{\partial \Omega} \vec{t} \cdot \delta \text{\textbf{u}} dA = 0\,,\label{eq:weakform}
\end{equation}
with $\ten{P}$ denoting the first Piola-Kirchoff stress, $\ten{F}$ the work-conjugate deformation gradient, $\vec{b}$ a volumetric load, $\vec{\text{\bf{u}}}$ the displacement, $\vec{t}$ a nominal traction on boundary $\partial \Omega$, and $dA$ and $dV$ infinitessimal surface and volume elements, respectively. $\delta \vec{\text{\bf{u}}}$ and $\delta \ten{F}$ refer to kinematically admissible variations in $\vec{\text{\bf{u}}}$ and $\ten{F}=\frac{\partial (\vec{X+\text{\bf{u}}})}{\partial \vec{X}}$, respectively, under which the variation of the virtual work must vanish~\cite[p.82]{Wri:2001:nf}.

A material-dependent constitutive relation $\ten{P}= \ten{P}(\ten{F})$ links strain and stress; in the hyperelastic case with which the current work is concerned, the first Piola-Kirchhoff stress is assumed to be derivable from a stored energy function $W$ as the first derivative $\ten{P} = \frac{\partial W}{\partial \ten{F}}$~\cite[p.207]{Hol:2002:nsm}. For the numerical examples in Section~\ref{s:results}, we consider a simple compressible neo-Hooke potential
\begin{equation}
    W = \frac{\mu}{2} ( I_c - 3 ) + \frac{\kappa}{4} ( J^2 - 1 - 2 \ln{J} )\,,\label{eq:nH}
\end{equation}
with $I_c=F_{\beta \alpha}F_{\beta \alpha}$, $J=\text{det}(\ten{F})$, and $\mu$ and $\kappa$ being the shear and bulk moduli.  
The fourth-order nominal stiffness tensor $\ten{\mathcal{A}}=\frac{\partial \ten{P}}{\partial \ten{F}}$ can then also be derived from this potential as the second derivative $\ten{\mathcal{A}}=\frac{\partial^2 W}{\partial\ten{F}^2}$.

The discretisation of Eq.~\eqref{eq:weakform} via a standard isoparametric Galerkin Ansatz with ${\text{\textbf{u}}}^e(\vec{X})=\sum_k N^{ek}(\vec{X}) \vec{u}^{ek}$ over a set of elements $e\in E$ yields 
\begin{equation}
    \sum_{e\in E} \delta \vec{u}^{ekT} \vec{g}^{ek}=0,\quad \text{with}\quad \vec{g}^{ek} = \int_{\Omega^e} \ten{P}\frac{\partial N^{ek}}{\partial \vec{X}} dV - \int_{\Omega^e} \vec{b} N^{ek} dV - \int_{\partial \Omega^e} \vec{t} N^{ek} dA \,,\label{eq:elemres}
\end{equation}
where $N^{ek}$ denotes the shape function corresponding to node $k$ in element $e$, with $\delta \vec{u}^{ek}$ being the corresponding nodal displacement and $\vec{g}^{ek}$ the associated nodal residual~\cite[p.101-125]{Wri:2001:nf}.
When all nodal variations $\delta {{u}}_\alpha^{ek}$ are collected in a vector $\delta \vec{u}$ without duplicating nodal degrees of freedom shared between elements, a global residual $\vec{g}$ can be defined. {$\vec{g}$ can be assembled with a complexity of roughly $\mathcal{O}(|E|\mathcal{C}_e^g)$ where $|E|$ is the total number of elements and $\mathcal{C}_e^g$ the cost of evaluating Eq.~\eqref{eq:elemres}, as} 
\begin{equation*}
    g_{3K+\alpha} = \sum_{e \in E} g_\alpha^{ek}\,,
\end{equation*}
where $K$ denotes the global node identifier associated with node $k$ in element $e$. Thus, the weak form becomes
\begin{equation}
    \delta \Vec{{u}}^T \Vec{g}(\Vec{{u}};\Vec{p}) = 0\,,\label{eq:variational}
\end{equation}
and since this must vanish for any kinematically admissible $\delta \vec{u}$,
\begin{equation}
    \Vec{g}(\Vec{{u}};\Vec{p}) = \vec{0}\,.\label{eq:eqsystem}
\end{equation}
Here, $\vec{p} \in \mathbb{R}^\delta$ denotes a vector of parameters which specifies an instance of the parametric problem class and implicitly determines a concrete solution $\vec{u}(\vec{p})$. $\vec{p}$ might for example specify the material constants $\kappa(\vec{X})$ and $\mu(\vec{X})$ or boundary conditions, e.g. via the traction $\vec{t}$ on boundary $\partial \Omega$.

For sufficiently well-behaved problems, roots of Eq.~\eqref{eq:eqsystem} can be sought using a classical Newton-Raphson scheme, in which case Eq.~\eqref{eq:eqsystem} is linearised around the current iterate $\vec{u}_\text{cur}$ and an increment $\Delta \vec{u}$ computed via~\cite[p.148]{Wri:2001:nf}
\begin{equation}
    \ten{K}(\vec{u_\text{cur}};\Vec{p}) \Delta {\Vec{u}} = - \Vec{g} (\vec{u_\text{cur}};\Vec{p})\,. \label{eq:newton}
\end{equation}
The Jacobian or stiffness matrix $\ten{K}$ appearing in the above is defined as
\begin{equation}
    \ten{K}(\vec{u_\text{cur}};\Vec{p}) = \frac{\partial \Vec{g}(\vec{u_\text{cur}};\Vec{p})}{\partial {\Vec{u}}}\,, \label{eq:K_glob}
\end{equation}
which can be assembled analogously to $\vec{g}$ at $\mathcal{O}(|E|\mathcal{C}_e^K)$, with the cost of computing the element stiffness $\mathcal{C}_e^K$, as
\begin{equation*}
    K_{3K+\alpha,3L+\beta} = \sum_{e \in E} K_{\alpha\beta}^{ekl}
\end{equation*}
in terms of an element stiffness matrix $\ten{K}^{ekl}$ (in index notation, using Einstein summation convention)
\begin{equation}
    {K}_{\alpha\gamma}^{ekl} = \int_{\Omega^e} \frac{\partial N^{el}}{\partial X_\beta} \mathcal{A}_{\alpha\beta\gamma\delta}  \frac{\partial N^{ek}}{\partial X_\delta} dV \,.\label{eq:K}
\end{equation}

\section{Computational homogenisation with RVEs}\label{s:RVE}

Computational homogenisation techniques such as the FE\textsuperscript{2} method model microstructural mechanical processes and their effect on macroscopic behaviour via a characteristic RVE. The evaluation of a material law on the macroscale $\ten{\Bar{P}}(\ten{\Bar{F}})$ is replaced by the solution of an RVE problem subject to suitable boundary conditions and the subsequent computation of the average stress and the associated stiffness. In the following, we review aspects of computational homogenisation which are important for our discussion of Galerkin- and hyperreduction later; please consult the reference literature for more detail~\cite{Sch:2014:nth,SaeSteJav:2016:ach}.
Below, macroscale quantities will be denoted by an overbar $\bar{\cdot}$, while microscale quantities remain undecorated.

The Hill-Mandel condition constitutes a physically sensible scale-coupling relation: the work done on a point $\vec{\bar{X}}$ on the macroscale via a strain perturbation $\Delta \ten{\bar{F}}$ must equal the work done on an associated microscale RVE via an associated perturbation $\Delta \vec{x}$ of the current position $\vec{x}(\vec{X})$ on the microscale, i.e.~\cite{Sch:2014:nth}
\begin{equation}
    \ten{\Bar{P}} : \Delta \ten{\Bar{F}} = \frac{1}{V} \int_{\partial \Omega} \vec{t} \cdot \Delta \vec{x} dA\,.\label{eq:hillmandel}
\end{equation}
Therein, the macroscopic deformation gradient $\ten{\bar{F}}$ and first Piola-Kirchhoff stress $\ten{\bar{P}}$ can be defined as
\begin{equation}
    \ten{\Bar{F}} = \frac{1}{V} \int_{\partial \Omega} \vec{x} \otimes \vec{N} dA\,,\quad \text{and} \quad  \ten{\Bar{P}} = \frac{1}{V} \int_{\Omega} \ten{P} dV\,,  \label{eq:FPmac}
\end{equation}
where $\vec{x}$ is the deformation on the microscale, assuming no tractions act on pores within the RVE~\cite{Sch:2000:hnk,Sch:2014:nth}.

In this work, we employ classical periodic boundary conditions, which satisfy Eq.~\eqref{eq:hillmandel} in a physically sensible way~\cite{Sch:2014:nth}. Thus, the displacement fluctuation
\begin{equation}
    \Tilde{\text{\textbf{u}}} = \vec{x} - \ten{\Bar{F}} \vec{X}\,, \label{eq:displfluc}
\end{equation}
is assumed to be periodic with the RVE acting as a unit cell. We do not discuss implementational details here, since boundary handling is not the focus of this work. The vector $\vec{\tilde{u}}\in \mathbb{R}^D$, which results from the discretisation of $\Tilde{\text{\textbf{u}}}$ and the application of boundary conditions, then becomes the primary unknown of the RVE problem.

To the accuracy of the FE discretisation, the macroscopic first Piola-Kirchhoff stress $\ten{\Bar{P}}$ can be postprocessed from the stress field on the RVE by volume averaging with around $\mathcal{O}(|E|\mathcal{C}_e^P)$, with $\mathcal{C}_e^P$ being the cost of computing the element-level stress contribution (see Eq.~\eqref{eq:FPmac})
\begin{equation}
    \ten{\bar{P}} = \frac{1}{V} \sum_e \int_{\Omega^e} \ten{P} dV\,. \label{eq:homogP}
\end{equation}
The macroscopic nominal stiffness $\ten{\mathcal{\bar{A}}}$ is defined as the rate of the change of $\ten{\bar{P}}$ with $\ten{\bar{F}}$, i.e. $\ten{\mathcal{\bar{A}}} = \frac{\partial \ten{\bar{P}}}{\partial \ten{\bar{F}}}$~\cite{Sch:2014:nth}. When written in Voigt notation and when the FE discretisation is substituted, this stiffness becomes
\begin{equation}
    \ten{\bar{A}}^v = \frac{1}{V} \sum_e \int_{\Omega^e} \ten{A}^v dv + \frac{1}{V} \ten{L}^T \frac{\partial \Delta \vec{\tilde{u}}}{\partial \Delta \vec{F}^v}\label{eq:homogA}\,,
\end{equation}
with the sensitivity coefficient $\ten{L}\in\mathbb{R}^{D\times 9}$, which can be assembled analogously to $\vec{g}(\vec{\tilde{u}};\ten{F})$ and $\ten{K}(\vec{\tilde{u}};\ten{F})$ from element contributions, in this case
\begin{equation}
    \mathcal{L}_{\epsilon \alpha \beta}^{ek} = \int_{\Omega^e} \mathcal{A}_{\alpha\beta\epsilon\zeta} \frac{\partial N^{ek}}{\partial X_\zeta} dV\,.\label{eq:L}
\end{equation}
For further details, please consult the reference literature, e.g.~\cite{Mie:2003:cmt,Sch:2014:nth}.

The first term in Eq.~\eqref{eq:homogA} can be assembled with around $\mathcal{O}(|E|\mathcal{C}_e^A)$, and and the degrees of freedom of $\ten{L}$ scale with about $\mathcal{O}(|E| \mathcal{C}_e^L)$, with $\mathcal{C}_e^A$ and $\mathcal{C}_e^L$ denoting the cost of the respective element-level operations.
The sensitivity $\frac{\partial \Delta \vec{\tilde{u}}}{\partial \Delta \vec{F}^v}$ appearing in the structural softening term can be computed as solutions to the variational problem
\begin{equation}
    \delta \vec{\tilde{u}}^T ( \ten{{K}} \Delta \vec{\tilde{u}} +\ten{{L}} \Delta \vec{\bar{F}}^v ) = 0\,, \quad \text{as} \quad \frac{\partial \Delta \vec{\tilde{u}}}{\partial \Delta \vec{F}^v} = \ten{S} = -\ten{K}^{-1} \ten{L}\,, \quad \text{and thus} \quad \ten{K} \ten{S} = \ten{L}\,.\label{eq:senssys}
\end{equation}
$\ten{S}\in \mathbb{R}^{D\times 9}$ is thus computed as the solution to 9 equation systems, where the matrix $\ten{K}$ is the same for all 9 cases, meaning that an appropriate sparse factorisation must be performed only once~\cite{Sch:2014:nth}. {The complexity of this step is of around $\mathcal{O}(D^2)$.}
The operations {and computational costs} involved in solving an RVE problem subject to periodic boundary conditions and homogenising the stress and stiffness are summarised in Alg.~\ref{alg:RVE_light}. For the sake of notational simplicity, no further distinction will be made between $\vec{u}$ and $\vec{\tilde{u}}$ in the following; the vector of the primary unknowns will be denoted as $\vec{u}$ for the sake of generality.

\begin{algorithm}[h!]
\caption{{Solve an RVE problem with the FEM}}
\While{$\|\ten{\bar{F}}-\ten{\bar{F}}_\text{target}\|>\text{tol}$}{
\tcp{Load increment}
$\ten{\bar{F}} \gets \ten{\bar{F}} +\Delta \ten{\bar{F}} $; \\
\While{$\text{max}(\text{abs}(\vec{g}))>\text{res}_{\text{max}}$}{
\tcp{Solve linearly for increment}
$\Delta {\Vec{{u}}} \gets \text{SOLVE} \left ( \ten{K} \Delta {\vec{{u}}} = - \vec{g} \right )$; \quad \tcp{around $ \mathcal{O}(D^2)$}
\tcp{Increment}
${\Vec{u}} \gets {\Vec{u}} + \Delta {\Vec{u}}$; \quad \tcp{$ \mathcal{O}(D)$}
\tcp{Assemble stiffness matrix and reaction vector}
$\ten{K},\vec{g} \gets \text{ASSEMBLE}$; \quad \tcp{$ \mathcal{O}(|E| \mathcal{C}_e), |E| \propto D$}
}
}
\tcp{Assemble sensitivity and apply boundary conditions}
$\ten{L}\gets \text{ASSEMBLE}$; \quad \tcp{$\mathcal{O}(|E| \mathcal{C}_e^L)$}
\tcp{Solve for homogenisation sensitivity}
$\ten{S} \gets \text{SOLVE}(\ten{K} \ten{S} = \ten{L})$; \quad \tcp{around $\mathcal{O}(D^2)$}
\tcp{Compute homogenised stress and Voigt stiffness}
$\ten{P} \gets \frac{1}{V} \sum_e \int_{\Omega^e} \ten{P}dV, \ten{\bar{A}}^v \gets \frac{1}{V} \sum_e \int_{\Omega^e} \ten{A}^v dV$; \quad \tcp{$\mathcal{O}(|E| \mathcal{C}_e^P)$ and $\mathcal{O}(|E| \mathcal{C}_e^A)$}
$\ten{\bar{A}} \gets \ten{\bar{A}}^v - \ten{L}^T \ten{S}$; \quad \tcp{$\mathcal{O}(D)$}
\label{alg:RVE_light}
\end{algorithm}

\section{Nonlinear Model Order Reduction}\label{s:NLMOR}

The computational cost of the iterative solver procedure outlined in Alg.~\ref{alg:RVE_light} is dominated by the assembly procedure with $\mathcal{O}(|E| \mathcal{C}_e)$, where $\mathcal{C}_e=\mathcal{C}_e^g+\mathcal{C}_e^K$ and the solution of the linear system with around $\mathcal{O}(D^2)$. The increments also scale with $\mathcal{O}(D)$, albeit with a small constant coefficient. As a practical contributor to computational cost, the number of iterations until convergence is also consequential, since runtime scales roughly linearly with it. The cost of post-convergence homogenisation, meanwhile, is dominated by the assembly of $\ten{L}$ with $\mathcal{O}(|E| \mathcal{C}_e^L)$, the solution of 9 linear problems with around $\mathcal{O}(D^2)$, stress and stiffness integration with $\mathcal{O}(|E| \mathcal{C}_e^P)$ and $\mathcal{O}(|E| \mathcal{C}_e^K)$, and (less significantly) the homogenised stiffness computation at $\mathcal{O}(D)$. The cost of element-level operations $\mathcal{C}_e^g, \mathcal{C}_e^K, \mathcal{C}_e^P$ and $\mathcal{C}_e^A$ are similar in magnitude, and, while vanishing compared to system-level operations, should not be underestimated: the evaluation of the constitutive relation, Ansatz functions, tensor operations, and numerical integration accumulate costs and runtime which are incurred for every element $e\in E$. Classical implementational considerations such as assembling $\vec{g}$ and $\ten{K}$ as well as integrating $\ten{P}$ and $\ten{A}^v$ in tandem mitigate this slightly, but not qualitatively.

In a parametric multi-query context -- e.g., computational homogenisation -- mitigating computational costs is crucial not only to accelerate investigations, but to make certain research feasible in the first place. As argued in the Introduction and in~\cite{SchFau:2024:mla}, projection-based MOR methods applied in combination with hyperreduction techniques strike an attractive balance between computational cost reductions, accuracy, data-economy, low offline cost, and generality in this circumstance. Broadly speaking, these approaches truncate the problem size by reducing the size of the solution space and the domain of integration. {If constructed well, this results in algorithms with a computational cost which no longer scales with the size of the original problem.} In the remainder of this Section, we discuss the first of the two major steps necessary to this end, while hyperreduction is covered in Section~\ref{s:HR}.

\subsection{Dimension reduction and representation learning problem}


The set of all solutions $\vec{u}$ to a parametric, quasi-static solid-mechanical problem defines a solution manifold $\mathcal{M}_{\vec{u}}=\{\Vec{{u}}\mid \exists \, \vec{p}: \vec{g}(\Vec{{u}};\Vec{p})=\Vec{0}\}$ for parameters $\Vec{p}\in \mathbb{R}^\delta$~\cite{DanCasAkk:2020:mor,DanCasAkk:2022:pca}.
If the problem is well-posed, this solution manifold is $\delta$-dimensional, meaning that solutions lie in a significantly lower-dimensional nonlinear subspace of the $D$-dimensional solution space. In the case of a hyperelastic RVE problem, for example, $\vec{p}$ contains the degrees of freedom of the macroscopic deformation gradient $\ten{\bar{F}}\in \mathbb{R}^{3\times 3}$ which are not responsible for rigid body rotations, such that $\delta = 6$~\cite{SchFau:2024:mla}. 
Projection-based MOR techniques attempt to exploit this observation by searching for solutions in a $d$-dimensional approximation space $\mathcal{M}_{\vec{\bar{u}}}$, where $d\ll D$. In the ideal case, $\mathcal{M}_{\vec{\bar{u}}}\supset \mathcal{M}_{\vec{u}}$, such that all possible solutions lie in $\mathcal{M}_{\vec{\bar{u}}}$. 

The solution manifold $\mathcal{M}_{\vec{u}}$ is of course not known a priori, such that projection-based MOR techniques have to have recourse to $s$ discrete snapshot solutions $\ten{U} = [ \Vec{u}_1,..,\Vec{u}_s] \in \mathbb{R}^{D\times s}, \Vec{u}_i \in \mathcal{M}_{\vec{u}}$ gathered from the underlying high-fidelity model in an offline training phase. 
The dimension reduction problem at the heart of nonlinear projection-based MOR could then be phrased as follows: given the snapshot data, find an embedding map $M:\mathbb{R}^D\rightarrow\mathbb{R}^d$ such that the embedding $\vec{y} = M(\vec{{u}})$ retains as much as possible of the structure of the solution manifold. Nonlinear projection-based MOR techniques can then seek solutions in the low-dimensional reduced space $\vec{y}\in \mathbb{R}^d$. To this end, a reconstruction map $R:\mathbb{R}^d\rightarrow\mathbb{R}^D$ defining approximate solutions $\vec{\bar{u}}=R(\vec{y})$ in the original solution space is also required, which turns the dimension reduction problem into a representation learning problem: find embedding and reconstruction maps $M:\mathbb{R}^D\rightarrow\mathbb{R}^d$ and $R:\mathbb{R}^d\rightarrow\mathbb{R}^D$, such that, for $d\ll D$, the reconstruction error $\sum_i \|\vec{\bar{u}}_i - \vec{u}_i\|=\sum_i \|R(M(\vec{u}_i))-\vec{u}_i\|$ is minimised. The set of reconstructions $\mathcal{M}_{\vec{\bar{u}}}=\{\vec{\bar{u}} \mid \vec{\bar{u}}=R(\vec{y}), \, \vec{y} \in \mathbb{R}^d \}$ then defines the aforementioned approximation space which is parameterised via the reduced variables $\vec{y}\in \mathbb{R}^d$.

\begin{figure*}[h!]
\centering
\begin{minipage}[t]{.45\textwidth}
  \centering
  \includegraphics[width=0.9\textwidth,,trim={8cm 4cm 7cm 3cm},clip]{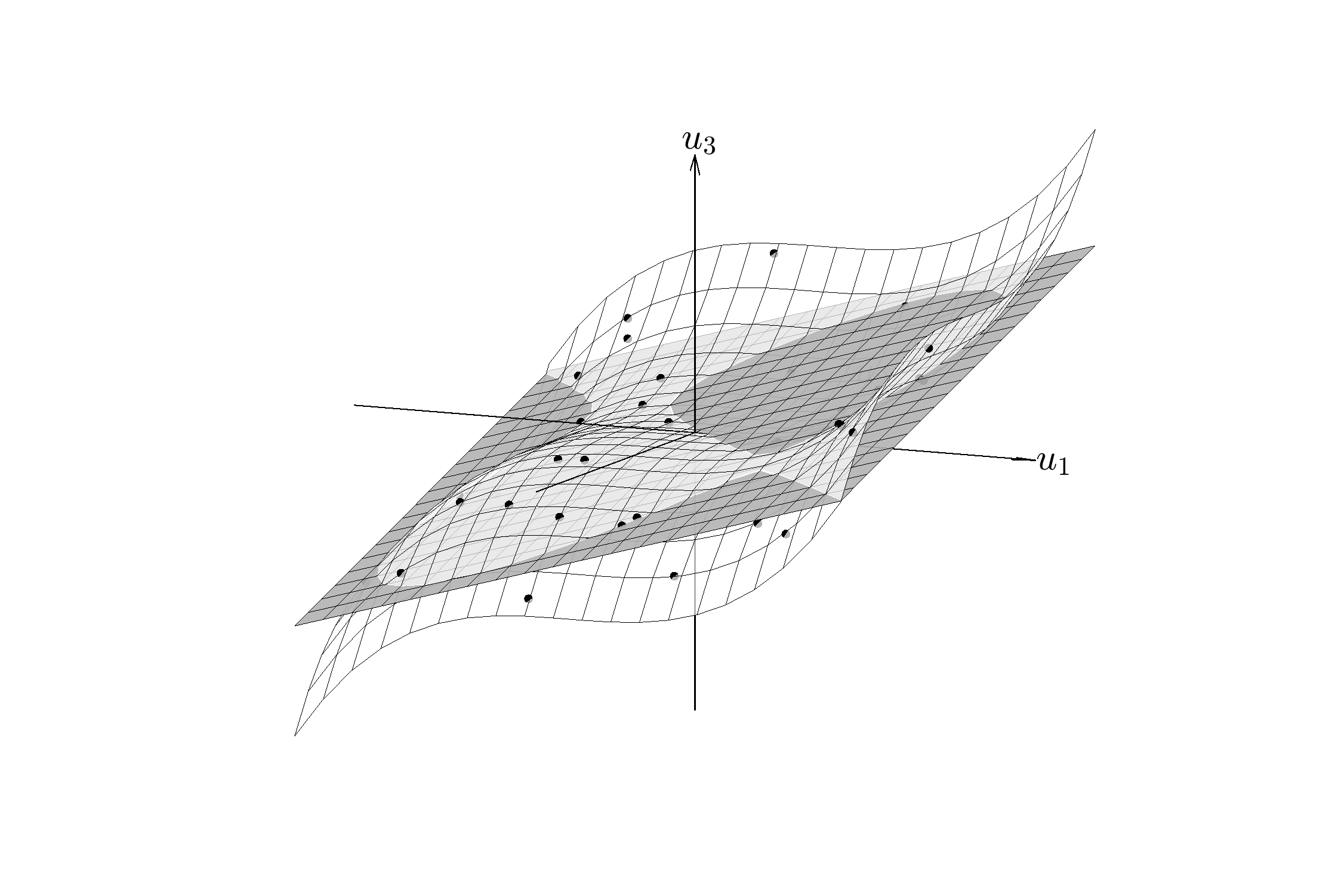}
  \caption{Illustration of a linear approximation space with $d=2$ obtained by the POD (gray plane) for a solution manifold with $\delta=2$ (meshed surface).}
\label{fig:POD}
\end{minipage}%
\hspace{0.5cm}
\begin{minipage}[t]{.45\textwidth}
  \centering
  \includegraphics[width=0.9\textwidth,,trim={8cm 4cm 7cm 3cm},clip]{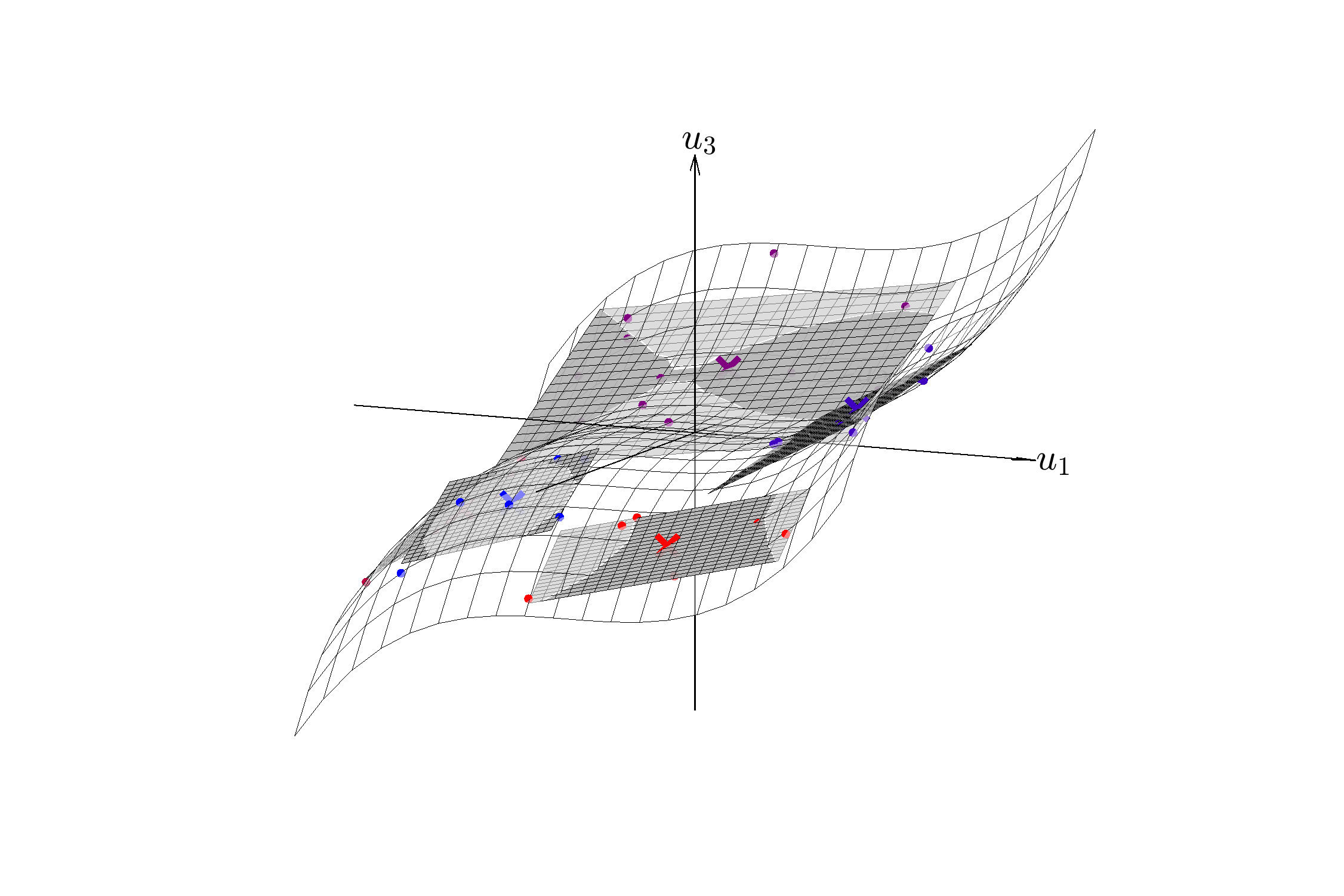}
  \caption{Illustration of a piecewise linear approximation space with $d=2$ obtained by the LPOD (gray planes) for a solution manifold with $\delta=2$ (meshed surface). Clusters of snapshots are illustrated as differently coloured dots, while $\times$-symbols denote the centroids of each cluster.}
\label{fig:LPOD}
\end{minipage}
\end{figure*}

\subsection{Approaches to defining approximation spaces}

In the following, we briefly cover several established approaches to defining approximation spaces, before recalling our manifold learning approach. For details, the interested reader is referred to the respective source material and to our previous work~\cite{SchFau:2024:mla}.

\subsubsection*{Proper Orthogonal Decomposition (POD)}

As motivated in the Introduction, the POD~\cite{Pea:1901:llp,Wei:2019:tpo,RadRee:2014:mre} is a popular approach to defining low-dimensional approximation spaces in the context of quasi-static solid mechanics. It defines linear embedding and reconstruction mappings
\begin{equation}
    {\Vec{y}} = \ten{{\psi}}^T \Vec{u}\,, \quad \text{and}\quad \vec{\bar{u}} = \ten{{\psi}} \Vec{y}\,,\label{eq:POD_map}
\end{equation}
where the mode matrix $\ten{\psi}$ can e.g. be obtained via a singular value decomposition $\ten{U} = \ten{L} \ten{\Sigma} \ten{R}$, as the leading $d$ left singular vectors $\vec{\psi}=[\vec{L}_1,..,\vec{L}_d]\in \mathbb{R}^{D\times d}$ of the snapshot matrix. Fig.~\ref{fig:POD} illustrates this approach on a low-dimensional data set.

\subsubsection*{Local Basis Method (LPOD)}

As noted in the Introduction, the POD might require comparatively high-dimensional reduced- $\vec{y}\in \mathbb{R}^d$ and approximation spaces $\mathcal{M}_{\vec{\bar{u}}}$ to accurately represent a highly nonlinear, $\delta$-dimensional solution manifold $\mathcal{M}_{\vec{u}}$. The LPOD~\cite{AmsZahFar:2012:nmo,AmsZahWas:2015:flr} aims to reduce the gap between $d$ and $\delta$ by localising the POD: snapshots $\vec{u}_i$ are clustered and local POD bases defined for each cluster. This conceptual approach is visualised in Fig.~\ref{fig:LPOD}. 
For the variant of the LPOD algorithm used here, as well as for parameter studies which were used to inform our choices for the (copious) model parameters, see~\cite{FauSch:2024:nmo}.
Despite its successes (for examples in quasi-static solid mechanics, see e.g.~\cite{GhaTisSim:2017:pdm,PagManQua:2018:nap,ChaAntBuf:2024:fpa}), the LPOD is subject to some limitations, especially in the data-poor regime. As we noted in previous work~\cite{SchFau:2024:mla,FauSch:2024:nmo}, in addition to the challenges of handling local basis transitions~\cite{IdeCar:1985:rmn,AmsZahWas:2015:flr,DanCasAkk:2020:mor,DanCasAkk:2022:pca,FauSch:2024:nmo}, choosing parameters properly and assuring cluster quality is nontrivial and locally linear approximations might not represent a solution manifold as closely as may be possible.

\subsubsection*{Polynomial manifold approach (PM)}

\begin{figure}[h]
    \centering
    \includegraphics[width=0.45\textwidth,trim={8cm 4cm 7cm 3cm},clip]{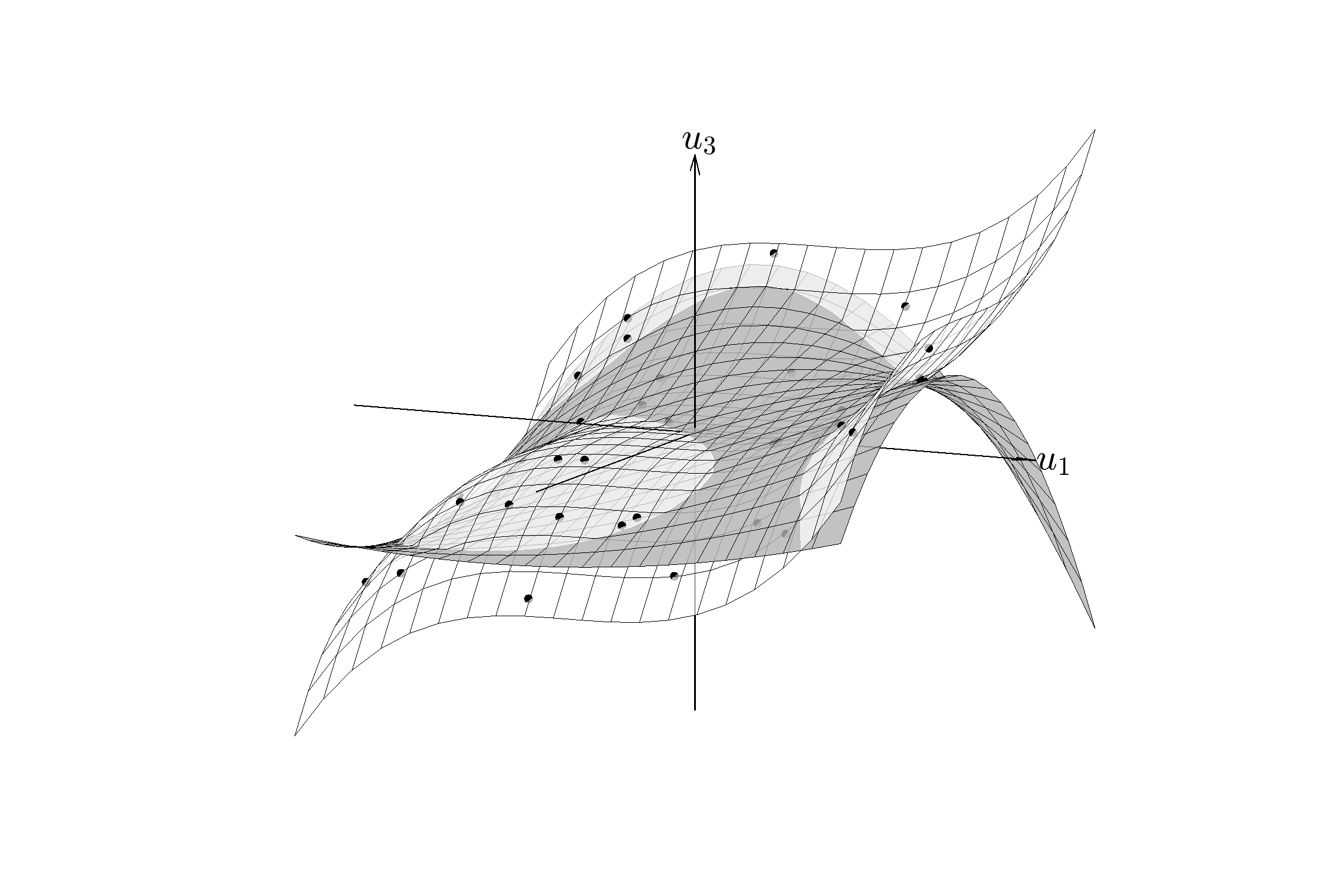}
    \caption{Illustration of polynomial manifold approximation space.}
    \label{fig:PM}
\end{figure}

The PM~\cite{JaiTisRut:2017:qmm,JaiTis:2019:hnm,GeeWriWil:2023:oin,BarFar:2022:qam} aims to avoid some of the shortcomings of the LPOD with a continuously nonlinear approximation space:
the reconstruction map is defined via a polynomial Ansatz
\begin{equation}
    \vec{\bar{u}} = R(\vec{y}) = \ten{\bar{V}} \ten{y} + \ten{\tilde{V}} \ten{\Xi} \vec{q}(\vec{y})\,,\label{eq:pm}
\end{equation}
where $\ten{\bar{V}}\in\mathbb{R}^{D\times d}$ and $\ten{\tilde{V}}\in\mathbb{R}^{D\times \tilde{d}}$ are orthonormal basis matrices.
$\vec{q}(\ten{y})\in\mathbb{R}^{p}$ contains polynomial terms of $\vec{y}$, e.g. monomials or full Kronecker products. Here, we use (vectorised) quadratic Kronecker products, i.e. $\vec{p}(\vec{y})=\text{vec}(\vec{y}\otimes \vec{y)}$.
$\ten{\Xi}\in \mathbb{R}^{\tilde{d}\times p}$ is a low-dimensional coefficient matrix for this nonlinear term, with $p$ being the number of polynomial terms. The coefficients $\ten{\bar{V}}, \ten{\tilde{V}}$, and $\ten{\Xi}$, as well as the embedding $\vec{y}_i$ of the snapshots $\vec{u}_i$ can be obtained by fitting the Ansatz to the snapshot data
\begin{equation*}
    \min_{\bm{\bar{V}},\bm{\tilde{V}},\bm{\Xi},\bm{y}_i}\sum_i\| \bm{u}_i-R(\bm{y}_i)\|^2\,,
\end{equation*}
e.g. via the alternating minimisation approach from~\cite{GeeWriWil:2023:oin}.

\subsection{Manifold learning approach}

\begin{figure}[h]
\centering
\begin{subfigure}[t]{.45\textwidth}
  \centering
  \includegraphics[width=0.95\textwidth,trim={8cm 4cm 7cm 3cm},clip]{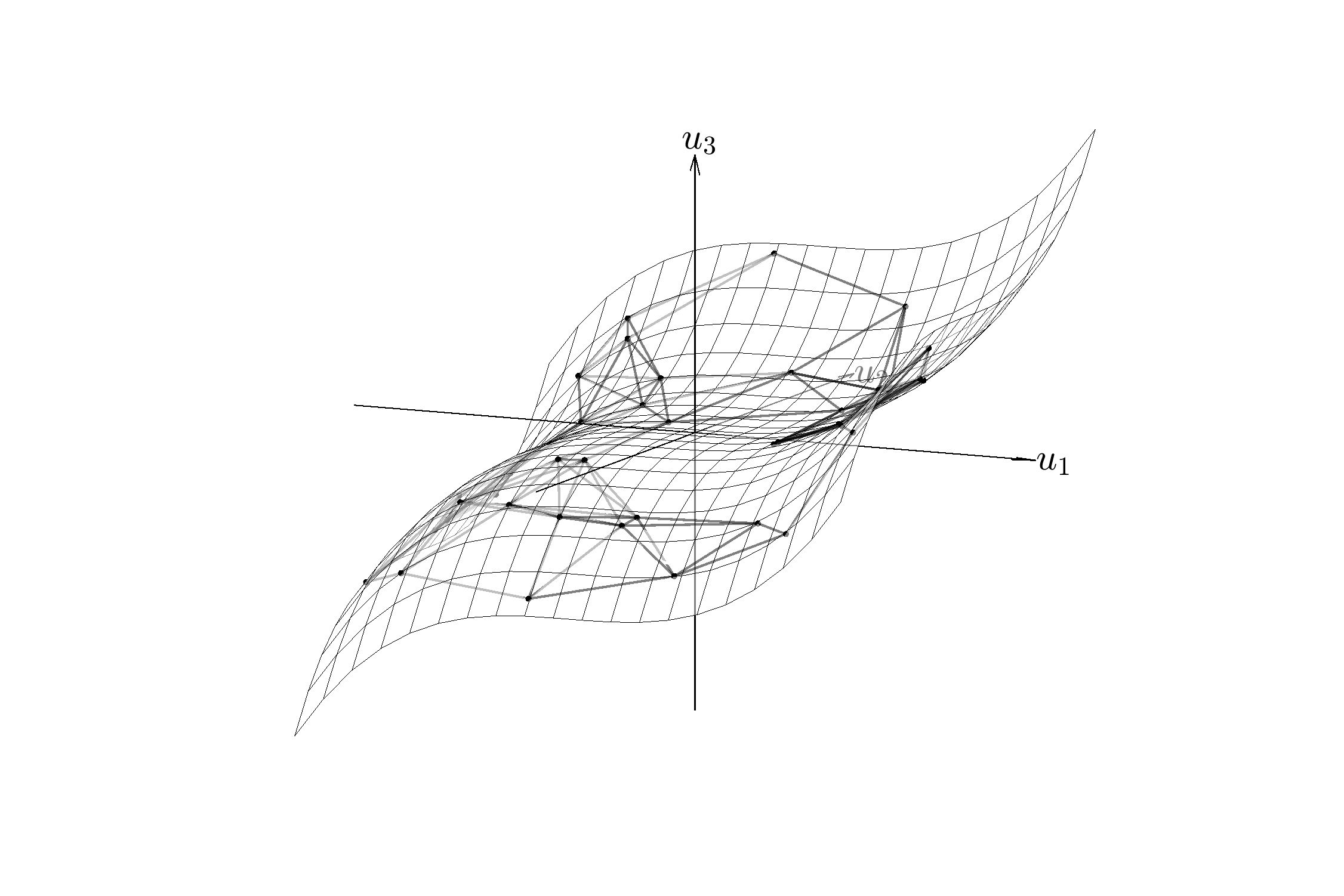}
  \caption{Illustration of k-nearest neighbour graph of snapshots (black dots), with edges being illustrated as black lines.}
  \label{fig:LLE2}
\end{subfigure}%
\hspace{0.5cm}
\begin{subfigure}[t]{.45\textwidth}
  \centering
  \includegraphics[width=0.9\textwidth,trim={8cm 3cm 4cm 2cm},clip]{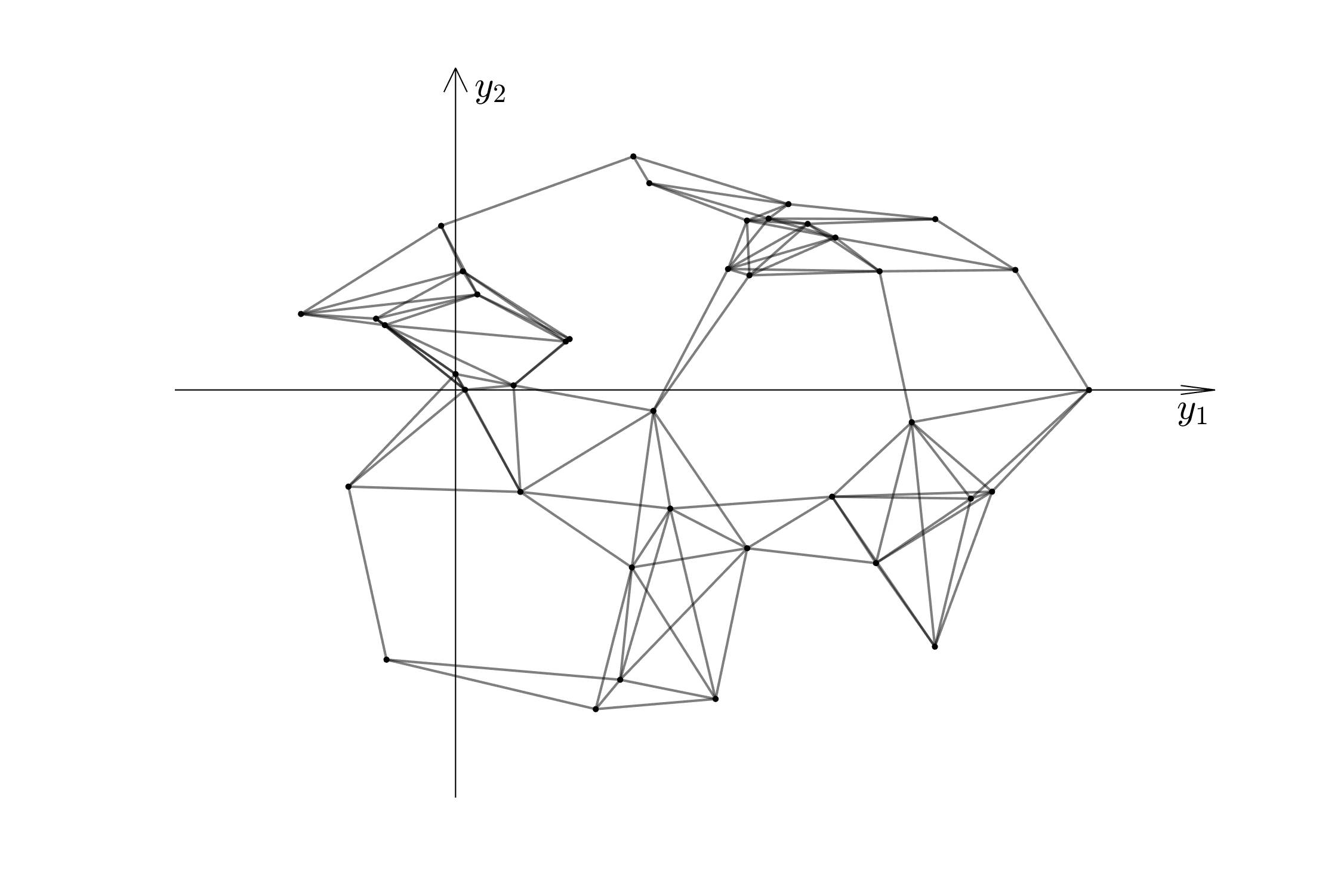}
  \caption{Illustration of embedding of snapshots in reduced space $\ten{Y} \in \mathbb{R}^{d\times s}$ {(black dots)} obtained via LLE.}
  \label{fig:LLE1}
\end{subfigure}%
\caption{Illustration of locally linear embedding.}
\label{fig:LLE}
\end{figure}

While the continuously nonlinear approximation space obtained by the PM is very attractive in the context of nonlinear MOR, the use of a specific polynomial Ansatz curtails the flexibility of this approach and to some extent limits the PM to solution manifolds which can be represented well globally by low-order polynomials. Of course, an ANN Ansatz could be substituted instead, but these are comparatively data-hungry and hence not suitable for the data-poor context in which we are interested here~\cite{JaiTisRut:2017:qmm,JaiTis:2019:hnm,GeeWriWil:2023:oin,BarFar:2022:qam}. 

In~\cite{SchFau:2024:mla}, we instead proposed a proof-of-concept for a graph-based manifold learning approach to the nonlinear MOR of quasi-static solid mechanical systems.
Such GML schemes work by constructing a graph adjacency matrix $\ten{G}\in \mathbb{B}^{s\times s}$ for the snapshot data $\ten{U} \in \mathbb{R}^{D\times s}$, e.g. via a k-nearest neighbours approach. Then, an embedding $\ten{Y}\in \mathbb{R}^{d\times s}$ which conserves essential structural characteristics of $\ten{U}$ and $\ten{G}$ is found. This embedding implicitly defines the reduced space $\mathbb{R}^d$ in which a projection-based MOR scheme can subsequently search for solutions. For a sufficiently well-behaved underlying (solution) manifold, it is often possible to successfully obtain $\ten{Y}$ based on only a few snapshots $s$, without recourse to a specific Ansatz~\cite{RowSau:2000:ndr,TenSilLan:2000:ggf,BelNiy:2003:led,ZhaZha:2004:pmn}. GML schemes thus combine the flexibility of an ANN approach with the data-economy of the POD.

Suitable GML schemes for our setting include Locally Linear Embedding (LLE)~\cite{RowSau:2000:ndr}, ISOMAP~\cite{TenSilLan:2000:ggf}, Laplacian Eigenmaps (LEM)~\cite{BelNiy:2003:led} and Local Tangent Space Alignment (LTSA)~\cite{ZhaZha:2004:pmn}. In this work, we exclusively employ LLE. After defining a local neighbourhood $N_i=\{j\mid G_{ij}=1\}$ of snapshot $i$, LLE constructs a reconstruction weight matrix $\ten{W}$ which optimally reconstructs each $\vec{u}_i$ from its neighbours $\vec{u}_j$, i.e.
\begin{equation*}
    \min_{\ten{W}} \sum_i \| \Vec{u}_i - \sum_{j \in N_i} W_{ij} \Vec{u}_j \|^2\,.
\end{equation*}
LLE then computes the embedding $\ten{Y}$ which optimally respects these reconstruction weights in the reduced space, i.e.
\begin{equation*}
    \min_{\ten{Y}} \sum_i \| \Vec{y}_i - \sum_{j \in N_i} W_{ij} \Vec{y}_j \|^2\,.
\end{equation*}
Note that solutions to these optimisation problems can be computed at low cost~\cite{RowSau:2000:ndr}. For further detail, the interested reader is referred to~\cite{RowSau:2000:ndr,SchFau:2024:mla}. An illustration of dimensionality reduction via the LLE can be found in Fig.~\ref{fig:LLE}.

\begin{figure}[h]
\centering
\begin{subfigure}[t]{.45\textwidth}
  \centering
  \includegraphics[width=0.9\textwidth,trim={8cm 3cm 4cm 2cm},clip]{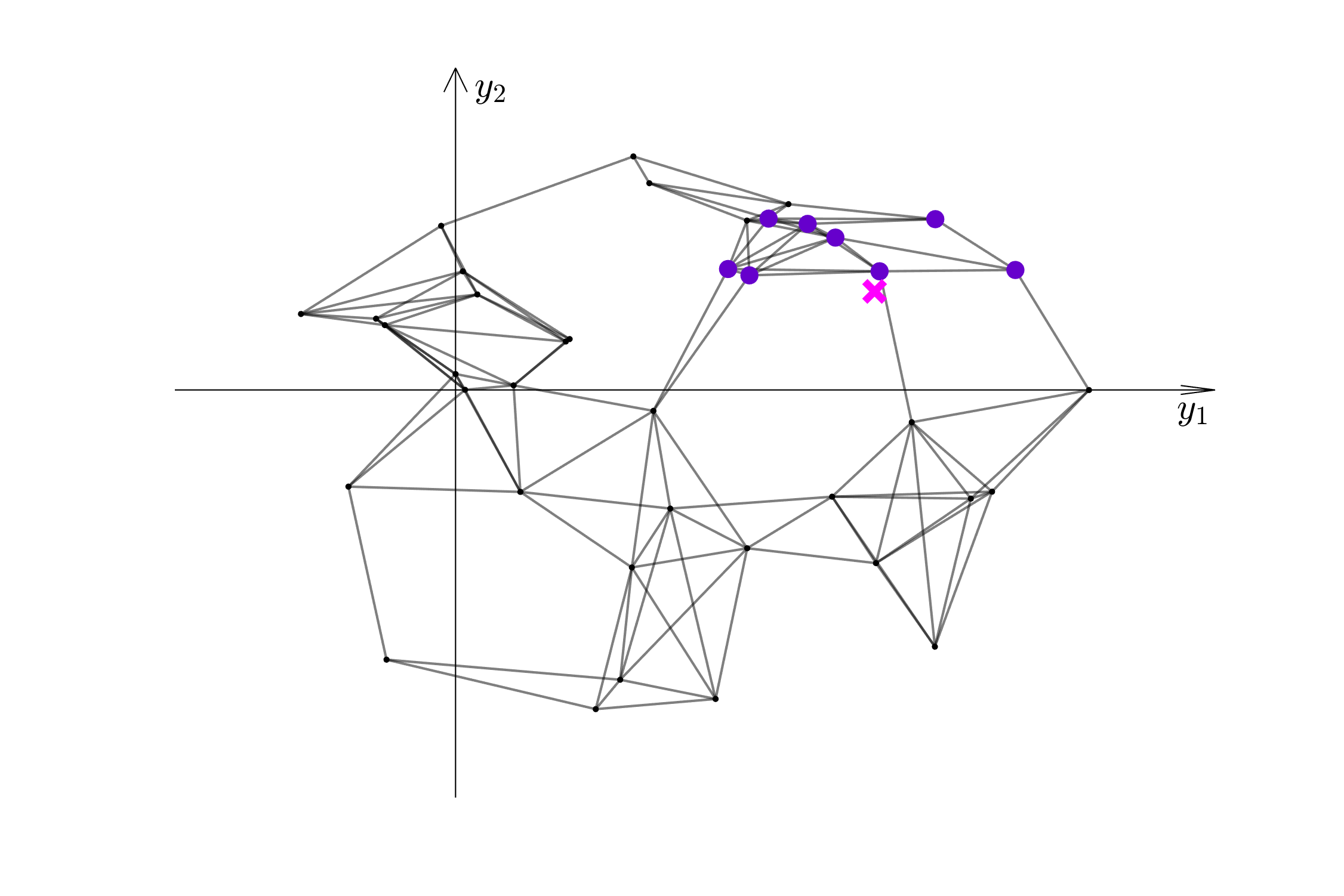}
  \caption{Illustration of neighbouring snapshots $\ten{Y}_n \in \mathbb{R}^{d\times n}$ {(purple dots)} of current iterate $\vec{y}$ {(magenta cross)} in reduced solution space $\Vec{y} \in \mathbb{R}^d$ obtained by LLE.}
  \label{fig:lin1}
\end{subfigure}%
\hspace{0.5cm}
\begin{subfigure}[t]{.45\textwidth}
  \centering
  \includegraphics[width=0.95\textwidth,trim={8cm 4cm 7cm 3cm},clip]{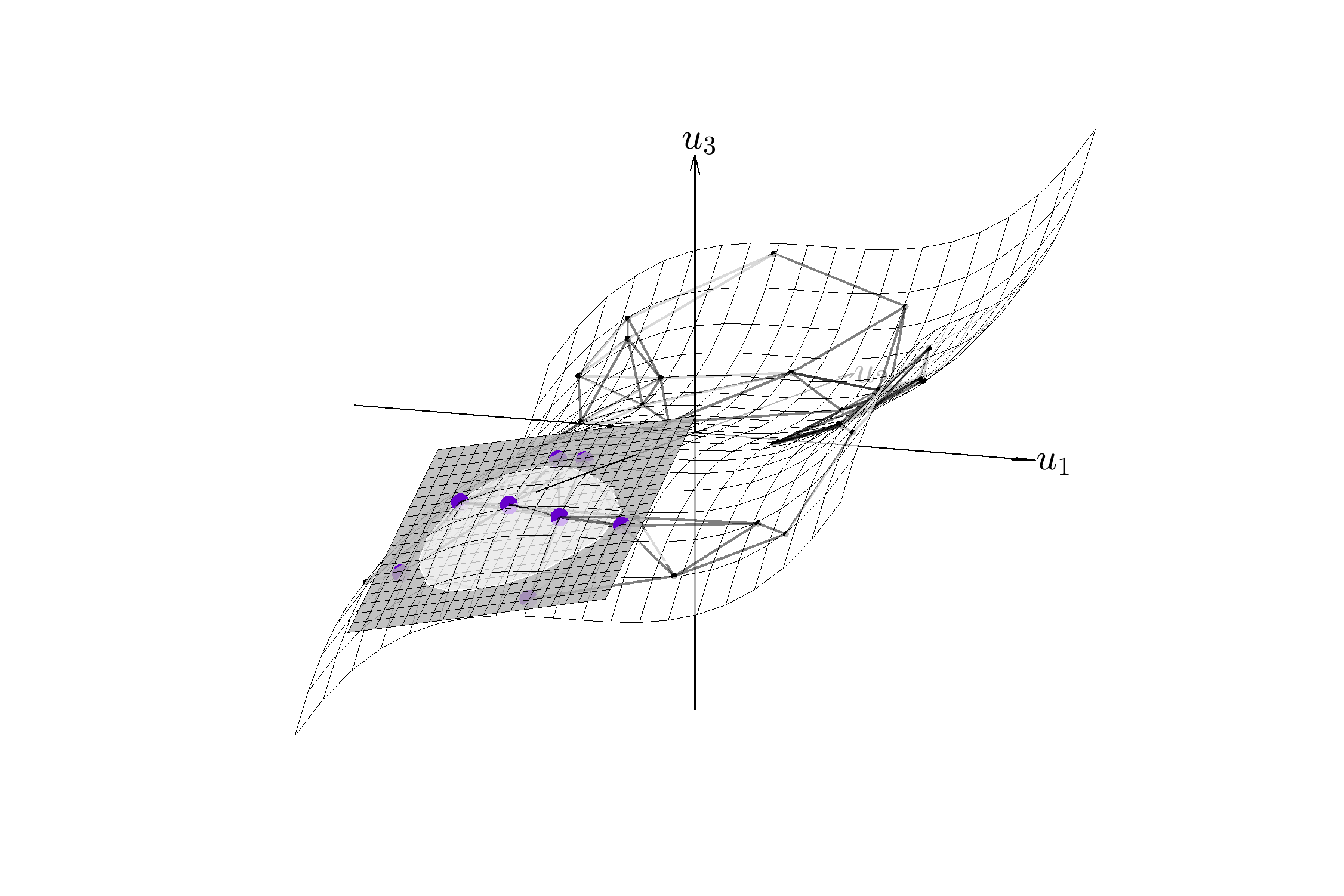}
  \caption{Illustration of approximate tangent to approximation space computed via neighbouring snapshot vectors {(purple dots)}. }
  \label{fig:lin2}
\end{subfigure}
\caption{Illustration of local linearisation.
}
\label{fig:lin}
\end{figure}

LLE achieves flexibility in the construction of the embedding at the cost of not obtaining explicit embedding $M:\mathbb{M}^D\rightarrow\mathbb{R}^d$ and reconstruction $R:\mathbb{R}^d\rightarrow\mathbb{R}^D$ maps. In~\cite{SchFau:2024:mla}, we obtained a reconstruction via a simple data-based local linearisation scheme deployed in the online phase. 
By linearising around the current value of the reduced variable $\vec{y}_\text{cur}$, we obtain a local model
\begin{equation}
    \Vec{\bar{u}} = \ten{\varphi} \Vec{y} + \Vec{u}^0\label{eq:loclin_param}\,,
\end{equation}
with tangent $\ten{\varphi}\in\mathbb{R}^{D\times d}$ based on the reduced $\Vec{y}_i \in \mathbb{R}^d, i \in N_y$ and original coordinates $\Vec{u}_i \in \mathbb{R}^D, i \in N_y$ of the $N$ snapshots $N_y$ nearest to ${\Vec{y}}_\text{cur}$. The neighbour search among $\vec{y}_i \in \mathbb{R}^d, i \in 1,..s$ can be computed at a low cost of $\mathcal{O}(sN)$.
Minimising the least squares error
\begin{equation*}
    \min_{\ten{\varphi},\Vec{u}^0} f(\ten{\varphi},\Vec{u}^0) = \min_{\ten{\varphi},\Vec{u}^0} \sum_n \sum_i ( \sum_j \varphi_{ij} Y_{Njn} + u_i^0 - U_{Nin} )^2 \,
\end{equation*}
yields
\begin{equation}
    \ten{\varphi} = \ten{U}_N \ten{W}_N \ten{Y}_N^T (\ten{Y}_N \ten{W}_N \ten{Y}_N^T)^{-1}\,.\label{eq:loclin_mat}\quad \text{with} \quad \ten{W}_N = \ten{I}_N - \frac{1}{N} \ten{1}_N\,.
\end{equation}
An illustration of the local linearisation scheme can be found in Fig.~\ref{fig:lin}.

In~\cite{SchFau:2024:mla}, we performed this linearisation in each Newton iteration in an Euler forward-like scheme.
Here, we instead note that at the beginning of each load step within an RVE computation, the macroscopic deformation gradient $\ten{\bar{F}}$ at the targeted solution is known (see Alg.~\ref{alg:RVE_light}). More generally, in a quasi-static solid mechanical problem, the parameters $\vec{p}$ determining the sought solution $\vec{u}$ are available at the beginning of each load step. Thus, we can perform the neighbour search among the snapshot parameter values $\ten{\bar{F}}_i, i\in 1,..,s$ rather than among the reduced snapshots $\vec{y}_i \in \mathbb{R}^d, i \in 1,..s$, and obtain an approximate Euler backward linearisation scheme. Unsurprisingly, this improves convergence and reduces runtime, since linearisations within the Newton scheme are avoided. We have also investigated more sophisticated strategies, such as an approximate Crank-Nicolson-like linearisation, but did not find these to be necessary in this work.

Finally, we deploy the embedding and reconstruction operations in a two-stage manner. Firstly, a linear compression to an intermediate, $\bar{d}$-dimensional space can be performed via the POD
\begin{equation}
    {\Vec{\bar{y}}} = \ten{\bar{\varphi}}^T \Vec{u}\,, \quad \text{and}\quad \vec{\bar{u}} = \ten{\bar{\varphi}} \Vec{\bar{y}}\,,
\end{equation}
followed by dimensionality reduction via the LLE into the reduced space $\mathbb{R}^d$. If $\bar{d}=s$, the first stage becomes a lossless compression into the span of the snapshots. Then, a local online linearisation can be performed to the intermediate space in the online phase at $\mathcal{O}(\bar{d}Nd)$ via
\begin{equation}
    \ten{\tilde{\varphi}} = \ten{\bar{Y}}_N \ten{W}_N \ten{Y}_N^T (\ten{Y}_N \ten{W}_N \ten{Y}_N^T)^{-1}\,,\label{eq:loclin_mat_two_stage}
\end{equation}
where $\ten{\bar{Y}}$ denotes snapshots in the intermediate space, after compression via the POD.
{A two-stage local linearisation with $\ten{\varphi} = \ten{\bar{\varphi}} \ten{\tilde{\varphi}}$
can thus be computed at computational cost which no longer scales with $\mathcal{O}(D)$, with no (or, if $d<\bar{d}<s$, little) additional error.}

\subsection{Galerkin-reduced solution procedure}

With the solution to the representation learning problem at hand and the nonlinear approximation space $\mathcal{M}_{\vec{\bar{u}}}$ defined, we can turn our attention to the search for solutions within this approximation space. We aim to search for solutions to Eq.~\eqref{eq:variational} in $\mathcal{M}_{\vec{\bar{u}}}$, i.e.
\begin{equation*}
    \delta \vec{\bar{u}}^T \vec{g}(\vec{\bar{u};\vec{p}}) = 0\,, \quad \text{subject to} \quad \vec{\bar{u}} \in \mathcal{M}_{\vec{\bar{u}}}\,, \delta \vec{\bar{u}} \in \mathcal{T}_{\vec{\bar{u}}} \mathcal{M}_{\vec{\bar{u}}}\,,
\end{equation*}
with the variation $\delta \vec{\bar{u}}$ in the tangent space of the approximation manifold.
With the approximation space parameterised in terms of $\vec{y}$ via the reconstruction $\vec{\bar{u}}= R(\vec{y})$, we have $\delta \vec{\bar{u}} = \frac{\partial R(\Vec{y})}{\partial \Vec{y}}\Bigr|_{\Vec{y}} \delta \vec{y}$. Approximating the tangent to the approximation space via a local linearisation, i.e. $\ten{\varphi} \approx\frac{\partial R(\Vec{y})}{\partial \Vec{y}}\Bigr|_{\Vec{y}}$, we obtain
\begin{equation*}
    \delta \vec{y}^T \ten{\varphi}^T\vec{g}(\vec{\bar{u}};\vec{p})=0\,, \quad \text{subject to} \quad \vec{\bar{u}} \in \mathcal{M}_{\vec{\bar{u}}}\,,
\end{equation*}
which defines a reduced equation system
\begin{equation}
    \vec{g}_{\text{red}}(\vec{\bar{u}};\vec{p}) =  \ten{\varphi}^T \vec{g}(\vec{\bar{u}};\vec{p})=\vec{0}\,, \quad \text{subject to} \quad \vec{\bar{u}} \in \mathcal{M}_{\vec{\bar{u}}}\label{eq:g_red}
\end{equation}
with the reduced residual $\vec{g}_{\text{red}}$.
As in the full-order system, we can use a Newton-Raphson scheme to seek roots of Eq.~\eqref{eq:g_red}, i.e.
\begin{equation}
    \ten{K}_{\text{red}}(\vec{\bar{u}}_{\text{cur}};\vec{p}) \Delta \vec{y} = - \vec{g}(\vec{\bar{u}}_{\text{cur}};\vec{p})\,.\label{eq:solve_red}
\end{equation}
The reduced Jacobian $\ten{K}_{\text{red}}$ can be obtained by linearisation as
\begin{equation}
    \ten{K}_\text{red}(\vec{\bar{u}};\vec{p}) = \frac{\partial \vec{g}_{\text{red}}}{\partial \vec{y}} = \frac{\partial \ten{\varphi}^T \vec{g}}{\partial \vec{u}} \frac{\partial \vec{u}}{\partial \vec{y}} = \ten{\varphi}^T \ten{K}(\vec{\bar{u}};\vec{p})\ten{\varphi}\,.\label{eq:K_red}
\end{equation}
As discussed in~\cite{SchFau:2024:mla}, the cost of solving the reduced equation system in Eq.~\eqref{eq:solve_red} now scales with $\mathcal{O}(d^3)$ instead of around $\mathcal{O}(D^2)$, which slashes computational cost significantly. The assembly of $\vec{g}$ and $\ten{K}$, however, still scales with $\mathcal{O}(|E| \mathcal{C}_e)$. Less significantly, the computation of $\vec{g}_{\text{red}}$ via Eq.~\eqref{eq:g_red} and $\ten{K}_{\text{red}}$ via Eq.~\eqref{eq:K_red} scale with $\mathcal{O}(D)$ if performed na\"ively. {The reduction of these costs using hyperreduction techniques is addressed in Sect.~\ref{s:HR}. Before this can be done, however, the Galerkin projection of the homogenisation procedure must be addressed.}

\subsection{Galerkin-reduced homogenisation}

The macroscopic algorithmically consistent stiffness can be computed via Eq.~\eqref{eq:homogA}. With the (nonlinear) Galerkin approximation, this becomes
\begin{align*}
    \ten{\bar{A}}^v &  = \frac{1}{V} \int \ten{A}^v dv + \frac{1}{V} \ten{L}^T \frac{\partial \Delta \vec{\bar{u}}}{\partial \Delta \vec{y}} \frac{\partial \Delta \vec{y}}{\partial \Delta \vec{F}^v} = \frac{1}{V} \int \ten{A}^v dv + \frac{1}{V} \ten{L}^T \ten{\varphi} \frac{\partial \Delta \vec{y}}{\partial \Delta \vec{F}^v}\nonumber \\
    & = \frac{1}{V} \int \ten{A}^v dv + \frac{1}{V} \ten{L}_{\text{red}}^T \frac{\partial \Delta \vec{y}}{\partial \Delta \vec{F}^v} \label{eq:stiffness_red}
\end{align*}
with $\ten{L}_{\text{red}} = \ten{\varphi}^T \ten{L}$.
The sensitivity $\ten{S}_{\text{red}} = \frac{\partial \Delta \vec{y}}{\partial \Delta \vec{F}^v}$, meanwhile, can be computed via Eq.~\eqref{eq:senssys}
\begin{align*}
    \delta \vec{\bar{u}}^T ( \ten{K} \Delta \vec{\bar{u}} + \ten{L} \Delta \vec{\bar{F}}^v ) & = 0 \\
    \delta \vec{y}^T \ten{\varphi}^T ( \ten{K} \ten{\varphi} \Delta \vec{y} + \ten{L} \Delta \vec{\bar{F}}^v ) & = 0 \\
    \delta \vec{y}^T ( \ten{K}_{\text{red}} \Delta \vec{y} + \ten{L}_{\text{red}} \Delta \vec{\bar{F}}^v ) & = 0
\end{align*}
such that
\begin{equation}
    \frac{\partial \Delta \vec{{y}}}{\partial \Delta \vec{F}^v} = \ten{K}_{\text{red}}^{-1} \ten{L}_{\text{red}}\label{eq:sensitivity_red}
\end{equation}
Again, the solution of the linear equation systems now scales with $\mathcal{O}(d^3)$ rather than around $\mathcal{O}(D^2)$. However, the assembly of $\ten{L}$ still scales with $\mathcal{O}(|E| \mathcal{C}_e^L)$ and the projection with $\mathcal{O}(D)$ if done na\"ively. The computation of $\ten{\bar{P}}$ and $\ten{\bar{A}}^v$ via Eq.~\eqref{eq:homogP} and Eq.~\eqref{eq:homogA} also still scale with $\mathcal{O}(|E| \mathcal{C}_e^P)$ and $\mathcal{O}(|E| \mathcal{C}_e^A)$, respectively. {Thus, an additional hyperreduction step also is required for these homogenisation operations.}

\section{Residual vector-based hyperreduction}\label{s:HR}

In simple terms, the Galerkin projection reduces the computational cost of solving the equation systems to which the Newton-Raphson solver and the homogenisation procedure give rise, but not the cost of evaluating constitutive relations and integrating local balance laws to assemble these equation systems.
As motivated in the Introduction, we further need to reduce the domain of integration via a second, hyperreduction step to achieve independence from the problem sizes $D$ and $|E|$ in the computational complexity and to achieve truly significant runtime reductions~\cite{Ryc:2009:hrm}.

In the context of the FEM, classical hyperreduction techniques such as the Gappy POD~\cite{EveSir:1995:klp,Wil:2006:ufls}, the Best Points Interpolation Method~\cite{NguPatPer:2008:bpi}, Missing Point Estimation~\cite{AstWeiWil:2008:mpe}, Reduced Integration Domain~\cite{Ryc:2009:hrm}, Discrete Empirical Interpolation Method (DEIM)~\cite{ChaSor:2010:nmr,TisDedRix:2013:dmd}, S-OPT~\cite{LauCheShi:2024:sps} Least Squares Petrov Galerkin~\cite{CarBouFar:2011:enl,CarBarAnt:2017:glp} or Reduced Over-Collocation~\cite{CheJiNar:2021:lrc}, Gauss-Newton with Approximate Tensors~\cite{CarFarCor:2013:gmn}, Energy Conserving Weighting and Sampling (ECSW)~\cite{FarAveCha:2014:drn}, Empirical Cubature~\cite{HerCaiFer:2017:dhn}, Statistically Compatible Hyperreduction~\cite{Wul:2024:sch}, and Empirically Corrected Cluster Cubature~\cite{WulHau:2025:ech} do this by reducing the effective number of elements $|E_m|<|E|$ or Gauss points which are considered for numerical integration. A low-order approximation can be successful because the reduced residual is a function of the location of the sought solution on the solution manifold $\vec{u} \in \mathcal{M}_{\vec{u}}$, as well as of the location of the current iterate in the approximation space $\vec{\bar{u}} \in \mathcal{M}_{\vec{\bar{u}}}$ -- and thus also lies on a low-dimensional manifold. Here, we consider only hyperreduction methods which accomplish this by assembling only few $m \ll D$ rows of the original nonlinear equation systems~\cite{Ryc:2009:hrm,EveSir:1995:klp,Wil:2006:ufls,NguPatPer:2008:bpi,AstWeiWil:2008:mpe,ChaSor:2010:nmr,TisDedRix:2013:dmd,LauCheShi:2024:sps,CarBouFar:2011:enl,CarFarCor:2013:gmn,CarBarAnt:2017:glp,CheJiNar:2021:lrc}. In keeping with existing literature in the context of the popular DEIM hyperreduction technique, we denote this set of degrees of freedom $P_m=\{m_1,...m_m\}$ of $\ten{K}$, $\vec{g}$, and $\ten{L}$ which are assembled exactly as {magic points}.

\subsection{Magic points selection}\label{s:magic}

The DEIM~\cite{ChaSor:2010:nmr} (for application to quasi-static solid mechanics, see also~\cite{RadRee:2016:pbm,FriHaaRyc:2018:ach}) uses a POD of residual snapshots to characterise the residual manifold via modes. Then, some characteristic entries of these modes (i.e., magic points) are selected via a greedy procedure. In this Subsection, we outline the magic point selection, which can also be employed in different hyperreduction schemes; the definition of the resulting hyperreduced equation systems are discussed in the following Subsections. 

To this end, $s_g$ residual snapshots $\ten{G}=[\vec{g}_1,..,\vec{g}_{s_g}]\in\mathbb{R}^{D\times s_g}$ are gathered from a Galerkin ROM. In contrast to the time-dependent case, in which physical dynamics are to be approximated, we later aim to characterise the ``dynamics" of a Newton-Raphson solver on the approximation space via the residual approximation. To faithfully represent the residual manifold, it is therefore paramount to collect residual snapshots from the intermediate, non-converged steps of the Newton-Raphson scheme. A POD of these residual snapshots $\ten{G}$ then yields $m$ modes $\ten{\Omega}\in \mathbb{R}^{D\times m}$ and mode coefficients $\vec{y}^{\vec{g}}\in \mathbb{R}^{m\times s_g}$ such that
\begin{equation*}
    \vec{g}_i \approx \ten{\Omega} \vec{y}_i^{\vec{g}}
\end{equation*}
with $\ten{\Omega} = [\vec{\omega}_1,\vec{\omega}_2,...,\vec{\omega}_m]$, where $m$ is the desired number of magic points.

Degrees of freedom which are predictive of these modes are selected iteratively. The first magic point is the maximal degree of freedom of the first POD mode
\begin{equation*}
    m_1 = \text{argmax}_i \omega_{1i}\,.
\end{equation*}
We also define a sampling matrix $\ten{Z}$, the first column of which is unit vector in $m_1$-th coordinate direction $\ten{Z} = [\vec{e}_{m_1}]$. 
For subsequent entries, we find the mode activity coefficients for the previous modes $\ten{\tilde{\Omega}}=[\vec{w}_1,..,\vec{\omega}_{j-1}]$ which best predict the magic points of the current mode
\begin{equation*}
    \ten{Z}^T \ten{\tilde{\Omega}} \vec{\tilde{y}^g} = \ten{Z}^T \vec{\omega}_j\,,
\end{equation*}
and subtract this contribution from the current mode
\begin{equation*}
    \tilde{\vec{\omega}}_j = \vec{\omega}_j - \ten{\tilde{\Omega}} \vec{\tilde{y}^g}\,,
\end{equation*}
to define the component $\tilde{\vec{\omega}}_j$ of the current mode $\vec{\omega}_j$ which is not predicted well by previous magic points. Subsequent magic points are the maximal degrees of freedom of this, as of yet unpredicted component
\begin{equation*}
    m_j = \text{argmax}_i \tilde{\omega}_{ji}\,,
\end{equation*} 
and the selection matrix is augmented with $\ten{Z} = [\ten{Z},\vec{e}_{m_j}]$.

The set of elements $E_m$ featuring magic points and the associated nodes and degrees of freedom $I_m$ define a reduced integration domain, via which the residual at the magic points $\vec{g}_m = \ten{Z}^T \vec{g}$
can be evaluated.
Then, the hyperreduced assembly of $\vec{g}_m$ scales roughly with $\mathcal{O}(|E_m|\mathcal{C}_e^g)$, rather than with $\mathcal{O}(|E|\mathcal{C}_e^g)$.

Meanwhile, the product of the magic rows of the stiffness matrix with the modes $\ten{\psi}$ can be computed on the element level. Denoting the product of $\ten{K}$ and $\ten{\psi}$ as $\widehat{\ten{K}\ten{\psi}}$, the assembly can be written as
\begin{equation}
    \widehat{K\psi}_{m_j k} = \sum_{e\in E_m} \sum_h k_{m_j^e h}^e \psi_{h,k}^e\,, \label{eq:elmprojk}
\end{equation}
where $m_j$ and $m_j^e$ denote the global and element indices of the magic points, respectively. $\ten{\psi}^e$ denotes the element-level projection matrix. This way, element stiffness matrices need only be computed for elements containing magic points, meaning that the hyperreduced assembly of $\widehat{\ten{K}_m\ten{\psi}}$ also scales roughly with $\mathcal{O}(|E_m|\mathcal{C}_e^K)$.
Furthermore, only the degrees of freedom of the $\ten{\psi}$ belonging to elements featuring magic points are required. 
Finally, only degrees of freedom of $\vec{u}$ belonging to elements in the reduced integration domain are required for the assembly, meaning that the required reconstruction also scales with $\mathcal{O}(|I_m|)$.

\subsection{Discrete Empirical Interpolation Method (DEIM)}\label{s:DEIM}

Once the magic points of the residual $\vec{g}_m$ have been assembled, the DEIM~\cite{ChaSor:2010:nmr} seeks to estimate the residual $\vec{g}$ via its modes $\ten{\Omega}$ and mode coefficients $\vec{y}^{\vec{g}}$. $\vec{y}^{\vec{g}}$ can be estimated via an interpolation constraint: when reconstructing $\vec{g}$ from $\vec{y}^{\vec{g}}$, the reconstructed values at the magic points should equal the known $\vec{g}_m$, i.e.
\begin{equation*}
    \vec{g}_m \overset{!}{=} \ten{Z}^T \ten{\Omega} \vec{y}^{\vec{g}}
\end{equation*}
The estimated mode coefficient therefore become
\begin{equation*}
    \vec{y}^{\vec{g}} = (\ten{Z}^T \ten{\Omega})^{-1} \vec{g}_m\,,
\end{equation*}
meaning that we can estimate residual and reduced residual as
\begin{equation}
    \vec{g} = \ten{\Omega} (\ten{Z}^T \ten{\Omega})^{-1} \vec{g}_m\,, \quad \text{and} \quad \vec{g}_{\text{hred}} = \ten{\psi}^T \ten{\Omega} (\ten{Z}^T \ten{\Omega})^{-1} \vec{g}_m\,.\label{eq:hrres}
\end{equation}
Consequently, the consistent hyperreduced stiffness matrix is given by
\begin{equation*}
    \ten{K}_{\text{hred}} = \frac{\partial \vec{g}_r}{\partial \vec{y}} = \ten{\psi}^T \ten{\Omega} (\ten{Z}^T \ten{\Omega})^{-1} \frac{\partial \vec{g}_m}{\partial \vec{u}} \ten{\psi} = \langle \ten{\psi}^T \ten{\Omega} (\ten{Z}^T \ten{\Omega})^{-1} \rangle \widehat{\ten{K}_m \ten{\psi}}\,.
\end{equation*}
In the linear case, the left factor of $\ten{K}_m$ and $\vec{g}_m$, $\ten{\psi}^T \ten{\Omega} (\ten{Z}^T \ten{\Omega})^{-1} = \ten{\psi}^T\ten{M} \in \mathbb{R}^{d\times m}$ can be preprocessed offline, such that the left projection only scales with $\mathcal{O}(m)$. In the nonlinear case, the product of $\ten{M}$ with the transpose of the first-stage linear compression matrix, $\ten{\bar{\varphi}}^T \ten{M} \in \mathbb{R}^{\bar{d}\times m}$, can similarly be preprocessed{, meaning that the DEIM allows us to achieve the desired reductions in computational costs.}

\subsection{Linear Extrapolation Hyperreduction Method (LEHM)}\label{s:LEHM}

We introduce a slight modification of the DEIM, which we call ``Linear Extrapolation Hyperreduction Method", i.e. LEHM, here, as this led to improvements in robustness in combination with nonlinear MOR techniques in the numerical experiments outlined below.
The DEIM reconstructs $\vec{g}$ via POD modes $\ten{\Omega}$ using an interpolation constraint. Instead, the reconstruction matrix $\ten{M}$ can also be computed more directly, via a least squares problem on the snapshot data $\vec{g}_j, j \in 1,..,s_g$, i.e.
\begin{equation*}
    \min_{\ten{M}} f(\ten{M}) = \min_{\ten{M}} \sum_j { \| \vec{g}_j-\ten{M} \vec{g}_j^m\|}\,.
\end{equation*}
The solution to this is given by 
\begin{equation}
    \ten{M} =  \ten{G} \ten{G}^{mT} (\ten{G}^m\ten{G}^{mT})^{-1}\label{eq:lehm}\,,
\end{equation}
where $\ten{G}^m$ denotes the rows of the snapshots matrix corresponding to magic points only.
Now, the residual and reduced residual can be approximated as
\begin{equation*}
    \vec{g} = \ten{M} \vec{g}^m\,, \quad \text{and} \quad 
    \vec{g}_{\text{hred}} = \ten{\varphi}^T \ten{M} \vec{g}^m\,.
\end{equation*}
Consequently, the reduced stiffness matrix is given by
\begin{equation*}
    \ten{K}_\text{hred} = \frac{\partial \vec{g}_{\text{red}}}{\partial \vec{y}} = \frac{\partial \ten{\varphi}^T \ten{M} \vec{g}^m}{\partial \vec{u}} \frac{\partial \vec{u}}{\partial \vec{y}} = \ten{\varphi}^T \ten{M} \widehat{\ten{K}^m  \ten{\varphi}}\,.
\end{equation*}
The linear system for the reduced increment, once approximated by extrapolation from the reduced integration domain, becomes
\begin{equation}
    \ten{K}_\text{hred} \Delta \vec{y} =\vec{g}_{\text{hred}} \label{eq:newtonhr}\,,
\end{equation}
which, just as in the case of the DEIM, can be assembled with significantly reduced computational complexity.

\subsection{Least-Squares Petrov Galerkin (LSPG)}\label{s:LSPG}

The DEIM and DEIM-like hyperreduction techniques assemble the magic points of the residual and estimate the remainder of the entries based on these via the reconstruction matrix $\ten{M}$. The hyperreduced residual is then computed by left projection onto the mode matrix $\ten{\varphi}$, meaning that these methods effectively approximate a Galerkin scheme~\cite{EveSir:1995:klp,Wil:2006:ufls,NguPatPer:2008:bpi,AstWeiWil:2008:mpe,ChaSor:2010:nmr,TisDedRix:2013:dmd,LauCheShi:2024:sps}. They do not, however, preserve the advantageous structure of such a Galerkin scheme (e.g. symmetries), which can lead to a loss of robustness~\cite{SolBraZab:2017:nsd,BraDavMer:2019:rmh}.

Instead, the LSPG and LSPG-like schemes bypass the approximate reconstruction of the full and the reduced residual entirely~\cite{CarBouFar:2011:enl,CarFarCor:2013:gmn,CarBarAnt:2017:glp,CheJiNar:2021:lrc}. Instead, they solve a reduced equation system defined on the magic points in a weighted or least-squares manner. 
The LSPG simply performs an overdetermined collocation at the magic points, iteratively solving the least squares problem~\cite{CarBouFar:2011:enl,CarFarCor:2013:gmn,CarBarAnt:2017:glp,CheJiNar:2021:lrc}
\begin{equation}
    \min_{\vec{y}}\|\widehat{\ten{K}_m  \ten{\varphi}}\Delta \vec{y}-\vec{g}_m\|\,.\label{eq:lspg}
\end{equation}
Standard solvers can be used to this end, with a computational cost of $\mathcal{O}(md^2)$. The resulting LSPG algorithm effectively approximates a left basis $\ten{K}\ten{\varphi}$ via a left projection with $\ten{\varphi}^T\ten{K}_m^T$ being applied to the magic points, hence justifying the classification as a Petrov-Galerkin scheme.

\subsection{Hyperreduced Homogenisation}

Recall that, as discussed in Subsections~\ref{s:DEIM} and~\ref{s:LEHM}, the non-magic rows of the residual and stiffness matrix can be reconstructed using a reconstruction matrix $\ten{M}$ in a DEIM-like scheme. Observing Eqs.~\eqref{eq:K} and~\eqref{eq:L}, we further note that the rows of $\ten{L}$ correlate with each other just like the rows of $\ten{K}$, meaning that $\ten{M}$ can also be used to reconstruct $\ten{L}$, i.e.
\begin{equation*}
    \ten{L} = \ten{M} \ten{L}^m\quad \text{and} \quad \ten{L}_{\text{hred}} = \ten{\varphi}^T \ten{M} \ten{L}^m\,.
\end{equation*}
Consequently 
\begin{equation*}
    \ten{\bar{A}}^v = \frac{1}{V} \int \ten{A}^v dv + \frac{1}{V} \ten{L}_{\text{hred}}^{mT}\ten{K}_{\text{hred}}^{-1} \ten{L}_{\text{hred}}\,,
\end{equation*}
or, for practical computation
\begin{equation*}
    \ten{K}_{\text{hred}} \ten{S}_{\text{hred}} = \ten{L}_{\text{hred}}\quad \text{and} \quad \ten{\bar{A}}^v = \frac{1}{V} \int \ten{A}^v dv + \frac{1}{V} \ten{L}_{\text{hred}}^{mT} \ten{S}_{\text{hred}}\,.
\end{equation*}
The inversions required for computational homogenisation can thus also be computed with $\mathcal{O}(d^3)$. 

However, the computation of the macroscopic stress via Eq.~\eqref{eq:homogP} and the Voigt term necessary in the algorithmically consistent tangent stiffness in Eq.~\eqref{eq:homogA} by integration scale with $\mathcal{O}(|E|\mathcal{C}_e^P)$ and $\mathcal{O}(|E|\mathcal{C}_e^A)$.
To reduce these computational cost, we may integrate only over elements in reduced integration domain $E_m$, and extrapolate from this reduced set of elements to the overall integral in an ECSW-like manner, i.e.
\begin{equation}
    \ten{\bar{P}}_{\text{hred}} = \sum_e \xi_e \ten{P}_e\,,
\end{equation}
with element coefficients ${\xi_e}$, at a cost of $\mathcal{O}(|E_m|\mathcal{C}_e^P)$~\cite{FarAveCha:2014:drn}. The vector of element coefficients $\vec{\xi}$ can be estimated from snapshots of the homogenised stress $\ten{\bar{P}}^{vh}\in\mathbb{R}^{9 s}$ and element stress $\ten{P}^{ve}\in\mathbb{R}^{9 s \times |E|}$. Here, the nine entries of stress snapshots in Voigt notation are stacked underneath each other.
To this end, we can minimise
\begin{equation*}
    \min_{\vec{\xi}} \| \ten{P}^{ve} \vec{\xi} - \ten{\bar{P}}^{vh} \|
\end{equation*}
via standard nonlinear nonnegative least squares solver. The resulting positive weights can be interpreted as larger effective relative volumes with which each element is multiplied; each element from the reduced integration domain thus additionally represents some share of the volume neglected in integration. The Voigt stiffness term can similarly be approximated via
\begin{equation*}
    \ten{\bar{A}}^v = \sum_e \xi_e \ten{A}^{ve} + \frac{1}{V} \ten{L}_{\text{hred}}^{mT} \ten{S}_{\text{hred}}\,,
\end{equation*}
such that its evaluations scales with $\mathcal{O}(|E_m|\mathcal{C}_e^A)$.

With this accelerated homogenisation procedure in place, we have now hyperreduced all computationally expensive operations in Alg.~\ref{alg:RVE_light}. The resulting solution- and homogenisation-algorithm, which is outlined as pseudocode in Algs.~\ref{alg:RVE_MOR_light_pre} and~\ref{alg:RVE_MOR_light}, does not feature operations with scale with the original problem size, i.e. with $\mathcal{O}(D)$ or $\mathcal{O}(|E|)$, any more.

\begin{algorithm}[h!]
\caption{{Preprocessing for hyperreduced nonlinear model order reduction}}
\tcp{Extract degrees of freedom of POD modes required for element-level projection}
$\ten{\bar{\varphi}}_m \gets \text{rows} \, i\in I_m \, \text{of} \, \ten{\bar{\varphi}}_{\text{full}}$\;
\tcp{Product of extended POD modes and residual reconstruction matrix for left factor}
$\ten{\bar{\varphi}}_M^T \gets \ten{\bar{\varphi}}_{\text{full}}^T \ten{M}$\;
\label{alg:RVE_MOR_light_pre}
\end{algorithm}

\begin{algorithm}[h!]
\caption{{Solve an RVE problem with DEIM-like hyperreduced nonlinear model order reduction}}
\While{$\|\ten{\bar{F}}-\ten{\bar{F}}_\text{target}\|>\text{tol}$}{
\tcp{Load increment}
$\ten{\bar{F}} \gets \ten{\bar{F}} +\Delta \ten{\bar{F}} $;\\
\tcp{Find nearest neighbours}
$\ten{Y}_N,\ten{U}_N \gets \text{NEIGH}$; \quad \tcp{$\mathcal{O}(9Ns)$}
\tcp{Compute linearisation}
$\ten{\tilde{\varphi}}\gets \ten{\bar{Y}}_N \ten{W}_N \ten{Y}_N^T (\ten{Y}_N \ten{W}_N \ten{Y}_N^T)^{-1}$; \tcp{$\mathcal{O}(\bar{d}Nd)$}
\While{$\text{max}(\text{abs}(\vec{g}))>\text{res}_{\text{max}}$}{
\tcp{Solve linearly for increment}
$\Delta {\Vec{y}} \gets \text{SOLVE} \left ( \ten{K}_{\text{hred}} \Delta {\vec{{y}}} = - \vec{g}_{\text{hred}} \right )$; \quad \tcp{$ \mathcal{O}(d^3)$}
\tcp{Compute increment of DOFs of unknown vector in reduced integration domain}
$\Delta \vec{u}_m \gets \ten{\bar{\varphi}}_m \ten{\tilde{\varphi}} \Delta y$; \quad \tcp{$\mathcal{O}(|I_m|\bar{d})$}
\tcp{Increment}
${\Vec{y}} \gets {\Vec{y}} + \Delta {\Vec{y}}$; \quad \tcp{$ \mathcal{O}(d)$}
${\Vec{u}_m} \gets {\Vec{u}_m} + \Delta {\Vec{u}_m}$; \quad \tcp{$ \mathcal{O}(|I_m|)$}
\tcp{Assemble magic points of residual and product of stiffness matrix and basis}
$\ten{K}_m^{\ten{\bar{\varphi}}},\vec{g}_m \gets \text{ASSEMBLE}$; \quad \tcp{$\mathcal{O}(|E_m|\mathcal{C}_e$)}
\tcp{Compute hyperreduced stiffness matrix and residual (reconstructed projection)}
$\ten{K}_{\text{hred}} \gets \ten{\tilde{\varphi}}^T \ten{\bar{\varphi}}_M^T \ten{K}_m^{\ten{\bar{\varphi}}} \ten{\tilde{\varphi}}$; \quad \tcp{$\mathcal{O}(\bar{d}m)$}
$\ten{g}_{\text{hred}} \gets \ten{\tilde{\varphi}}^T \ten{\bar{\varphi}}_M^T \vec{g}_m$; \quad \tcp{$\mathcal{O}(\bar{d}m)$}
}
}
\tcp{Assemble magic points of sensitivity coefficient}
$\ten{L}_m \gets \text{ASSEMBLE}$; \quad \tcp{$\mathcal{O}(|E_m|\mathcal{C}_e$)}
\tcp{Compute hyperreduced sensitivity coefficient}
$\ten{L}_{\text{hred}} \gets \ten{\tilde{\varphi}}^T \ten{\bar{\varphi}}_{\ten{M}}^T \ten{L}_m$; \quad \tcp{$\mathcal{O}(\bar{d}m)$}
\tcp{Solve for hyperreduced homogenisation sensitivity}
$\ten{S}_{\text{hred}} \gets \text{SOLVE}(\ten{K}_{\text{hred}} \ten{S}_{\text{hred}} = \ten{L}_{\text{hred}})$; \quad \tcp{$\mathcal{O}(d^3)$}
\tcp{Compute hyperreduced homogenised stress and Voigt stiffness}
$\ten{P} \gets \frac{1}{V} \sum_{e\in E_m} \xi_e \int_{\Omega^e} \ten{P}dV, \ten{\bar{A}}^{vv} \gets \frac{1}{V} \sum_{e\in E_m} \xi_e \int_{\Omega^e} \ten{A}^v dV$; \quad \tcp{$\mathcal{O}(|E_m| \mathcal{C}_e)$}
\tcp{Compute hyperreduced homogenised stiffness}
$\ten{\bar{A}}^v \gets \ten{\bar{A}}^{vv} - \ten{L}_\text{hred}^T \ten{S}_\text{hred}$; \quad \tcp{$\mathcal{O}(d)$}
\label{alg:RVE_MOR_light}
\end{algorithm}

\section{Numerical experiments}\label{s:results}

\begin{figure}[h]
    \centering
    \includegraphics[width=0.7\textwidth]{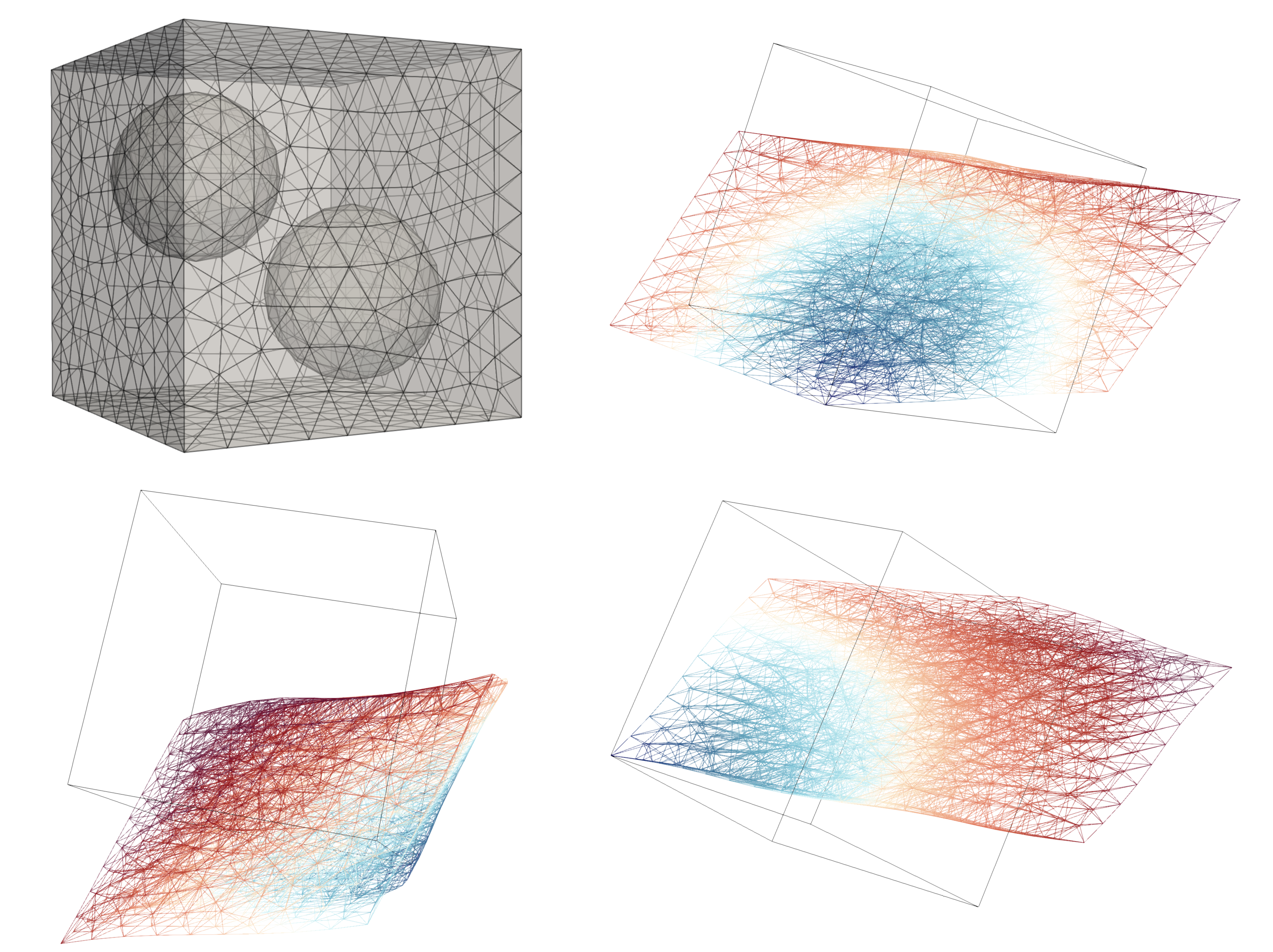}
    \caption{Discretised RVE with two inclusions alongside three example deformation states, shown to scale.}
    \label{fig:RVE}
\end{figure}

In this Section, we apply the MOR methods discussed above to a simple hyperelastic homogenisation problem. To this end, we consider an artificial RVE with two inclusions, as illustrated in Fig.~\ref{fig:RVE}. Both the matrix and the inclusions are modeled via the neo-Hookean stored energy function in Eq.~\eqref{eq:nH}, with Young's moduli of $E=1000$ for the matrix and $E=3000$ for the inclusions and Poisson's ratios of $\nu=0.2$ for both. As summarised in Tab.~\ref{tab:general_params}, the RVE has an edge length of $6$, while origin of the RVE coordinate system lies in a corner, such that $x_\text{min}=y_\text{min}=z_\text{min}=0$ and $x_\text{max}=y_\text{max}=z_\text{max}=6$. The inclusions each have a radius of $1.5$ and are centered at $(2,2,2)$ and $(4,4,4)$ in the RVE coordinate system, respectively. The microstructure and material behaviour result in a moderately nonlinear behaviour of the RVE. The microstructure is discretised using $4,821$ quadratic tetrahedral elements, resulting in $7,650$ nodes and $19,182$ independent degrees of freedom once boundary conditions are applied. Tab.~\ref{tab:general_params} summarises further problem parameters.

\begin{table}[h!]
    \centering
    \begin{tabular}{ c c c }
        \hline
        parameter & variable & value  \\
        \hline
        Young's modulus matrix & $E$ & 1000 Nmm\textsuperscript{-2} \\
        Young's modulus inclusions & $E$ & 3000 Nmm\textsuperscript{-2} \\
        Poisson's ratio & $\nu$ & 0.2 \\
        RVE edge length & & 6 mm \\
        inclusion radius & & 1.5 mm \\
        centre inclusion 1 ($x,y,z$) &  & (2,2,2) mm \\
        centre inclusion 2 ($x,y,z$) &  & (4,4,4) mm \\
        snapshot number & $s$ & 200 \\
        validation set size & & 500 \\
        step size & $\Delta {H}_{\text{LP}}$ & 0.03 \\
        perturbation size & $\Delta {H}_{\text{LS}}$ & 0.015 \\
        independent DOFs & $D$ & 19,182  \\ 
        element number & $n_\text{elem}$ & 4,821\\
        node number & $n_\text{node}$ & 7,650 \\
        \hline
    \end{tabular}    
    \caption{Parameters for the numerical experiments.}
    \label{tab:general_params}
\end{table}

\begin{figure}[h]
    \centering
    \includegraphics[width=0.45\textwidth,trim={8cm 3cm 8cm 2.5cm},clip]{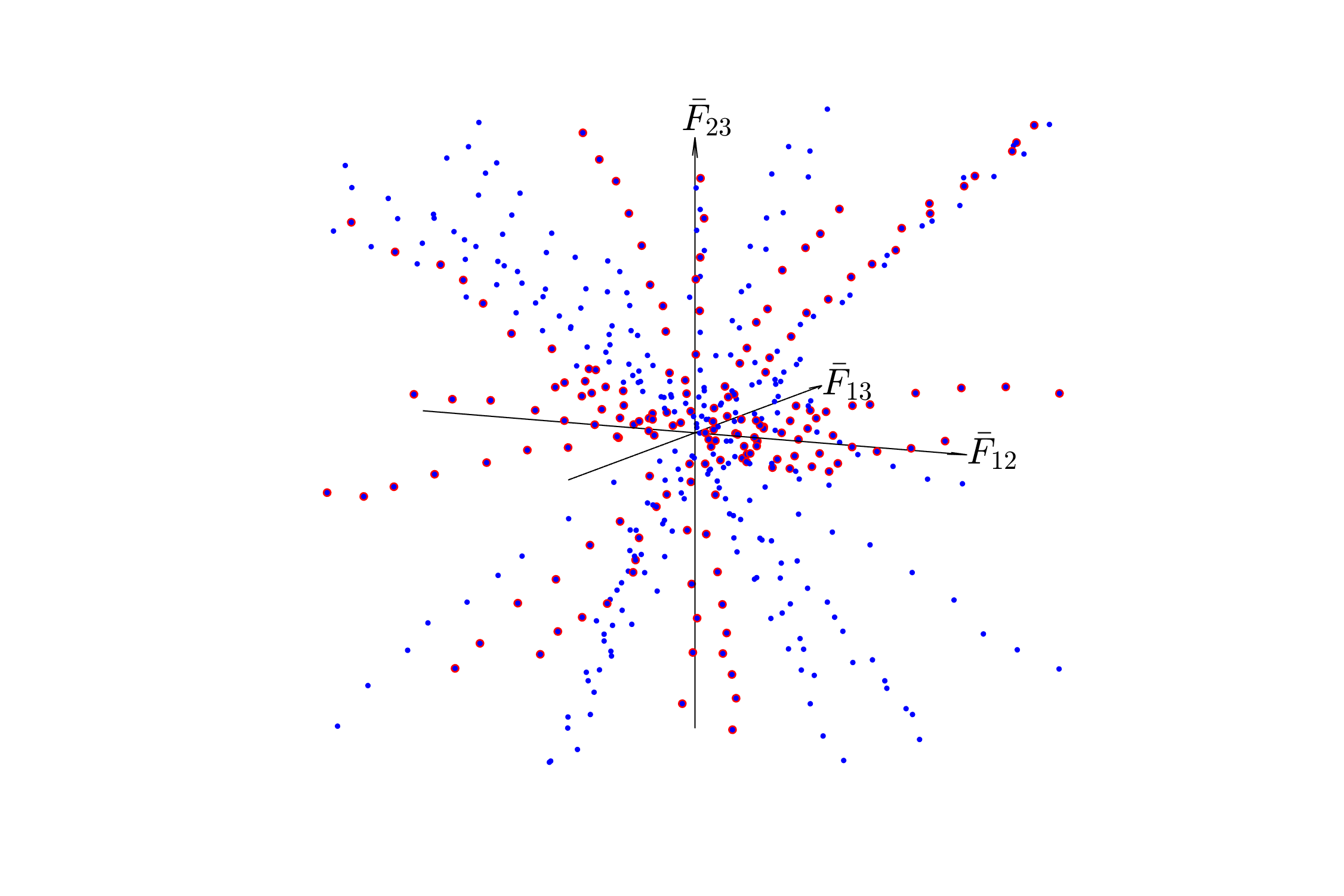}
    \caption{Three entries of macroscopic deformation gradient $\ten{\bar{F}}$ for training (red) and validation (blue) load paths.}
    \label{fig:RVE_loadpaths}
\end{figure}

Periodic boundary conditions are applied as outlined in Section~\ref{s:RVE}, with the macroscopic deformation gradient $\ten{\bar{F}}$ being prescribed. We define load paths consisting of $10$ load steps each: each load step consists of a step with length $\Delta {\bar{F}}_{\text{LP}}=0.03$ along a randomly sampled load direction $\ten{N}_{\text{LP}}$ which remains constant along the whole load path, as well as a perturbation of $\Delta {\bar{F}}_{\text{LS}}=0.015$ along a direction $\ten{N}_{\text{LS}}$ which is sampled again for each step, i.e.
\begin{align*}
    &\Delta \ten{\bar{F}} = \Delta {\bar{F}}_{\text{LP}} \ten{N}_{\text{LP}} + \Delta {\bar{F}}_{\text{LS}} \ten{N}_{\text{LS}}\,, \\&  \text{and} \quad
    \ten{\bar{F}} \leftarrow \ten{\bar{F}} + \Delta\ten{\bar{F}}\,.
\end{align*}
In Fig.~\ref{fig:RVE_loadpaths}, we illustrate three components of the macroscopic deformation gradients $\ten{\bar{F}}$ generated this way. Additionally, Fig.~\ref{fig:RVE} features deformation states of the RVE at the end of three example load paths, shown to scale.

\begin{figure*}[h!]
\centering
\begin{minipage}[t]{.45\textwidth}
    \centering
    \includegraphics[scale=1.0]{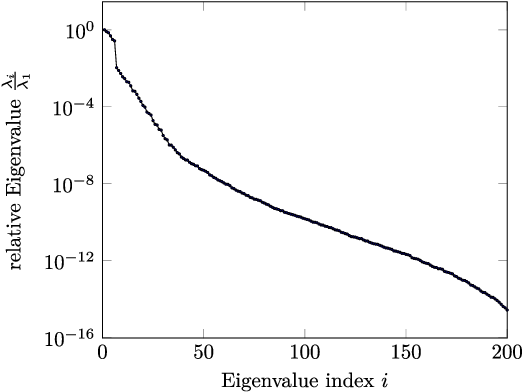}
    \captionof{figure}{Eigenvalue decay of the correlation matrix of the RVE snapshot data.}
    \label{fig:EV_decay}
\end{minipage}%
\hspace{0.5cm}
\begin{minipage}[t]{.45\textwidth}
    \centering
    \includegraphics[scale=1.0]{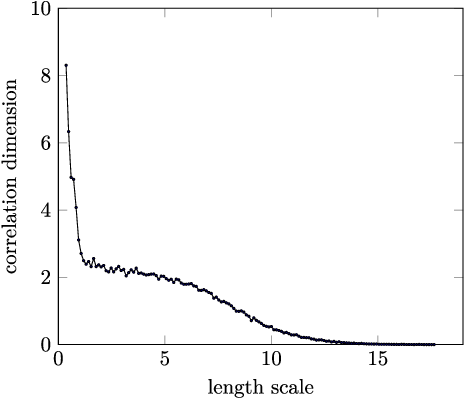}
    \captionof{figure}{Scale-dependent correlation dimension.}
    \label{fig:CD}
\end{minipage}%
\end{figure*}

For the numerical investigations, we first generate $s=200$ snapshot solutions $\ten{U}\in \mathbb{R}^{D\times s}$ along $20$ load paths via full FEM simulations. These training load paths are highlighted as red points in Fig.~\ref{fig:RVE_loadpaths}. The gradual Eigenvalue decay of the snapshot data correlation matrix, which is shown in Fig.~\ref{fig:EV_decay}, indicates that the snapshot data, and thus the solution manifold do not lie in a low-dimensional linear subspace, since otherwise, a rapid drop to zero would be observed. Meanwhile, the scale-dependent correlation dimension which is shown in Fig.~\ref{fig:CD} indicates empirically that the snapshots solutions indeed lie on a low-dimensional manifold of around $\delta=6$, since this measure approaches a value of around $8$ in the low length-scale limit.

Solution snapshots are then used to train nonlinear Galerkin ROMs using the techniques outlined in Section~\ref{s:NLMOR}. 
Then, the resulting Galerkin ROMs are used to compute solutions along the same $20$ load paths to obtain residual snapshots $\ten{G} \in \mathbb{R}^{D\times s_g}$ at the intermediate, i.e. non-converged, steps of the Newton-Raphson solver scheme, such that $s_g>s$ (alternatively, residual snapshots gathered from the full FEM model could be projected onto the relevant approximation spaces post-hoc). This residual training data is used to train hyperreduced models using the techniques outlined in Section~\ref{s:HR}.

Finally, the hyperreduced nonlinear ROMs are used to compute approximate solutions $\vec{\bar{u}}_i^{\text{ROM}}$ along the $30$ validation load paths (alongside the $20$ training load paths). Validation solutions $\vec{u}_i^\text{VAL}$ are obtained using a full FEM model for reference. Then, the mean error in the displacement field with respect to these validation solutions
\begin{equation}
    \text{Error} = \frac{1}{500} \sum_{i=1}^{500} \frac{\|\Vec{\bar{u}}_i^\text{ROM}-\vec{u}_i^\text{VAL}\|}{\|\Vec{u}_i^\text{VAL}\|}\,,\label{eq:error}
\end{equation}
is computed as a measure of ROM solver accuracy. Additionally, an equivalent error measure in the homogenised stress $\ten{\bar{P}}$ is computed as an indicator of ROM homogenisation accuracy. 

The performance of all MOR techniques investigated in this work is of course subject to the choice of several algorithmic parameters. The most important of these are the reduced model size $d$ -- i.e. the size of the reduced space -- and the hyperreduced model size $m$ -- i.e. the number of rows of $\vec{g}$ which are assembled exactly. We investigate the performance, measured in terms of accuracy, online speedup, and robustness, obtained using the MOR methods outlined in Sections~\ref{s:NLMOR} and~\ref{s:HR} for a range of values of $d$ and $m$ in the following Subsection. We do not investigate the influence of further parameters here. For all methods, these were chosen in view of preliminary parameter studies such as those communicated in~\cite{FauSch:2024:nmo,SchFau:2024:mla}. Parameters to which all methods are subject, such as those of the Newton-Raphson solver scheme, were of course chosen to be identical for all methods. 
All numerical investigations in this work were performed using an in-house python FEM and MOR library. 

\subsection{Results}

Tabs.~\ref{tab:POD_DEIM}, \ref{tab:POD_LEHM}, \ref{tab:PM}, and~\ref{tab:LLE} present the mean error 
over all $500$ solutions obtained using the POD, PM, and LLE in combination with DEIM-like hyperreduction techniques, for $d\in[9,3], m\in[20,250]$. High values of $E_{\text{mean}}>1\%$ are highlighted in light gray. Parameter combinations for which at least one of the $500$ simulations failed to converge are marked by a $\times$-symbol and highlighted in dark gray; no mean error was calculated in this case.

\begin{table}[h]
    \centering
    \begin{tblr}{
    colspec={cc|cccccccccc},
    cell{3}{3} = {gray},
    cell{3}{4} = {gray},
    cell{3}{8} = {gray},
    cell{3}{9} = {gray},
    cell{4}{4} = {gray},
    cell{4}{5} = {gray},
    cell{5}{5} = {gray},
    cell{5}{6} = {gray},
    cell{5}{7} = {gray},
    cell{6}{6} = {gray},
    cell{6}{7} = {gray},
    cell{6}{8} = {gray},
    cell{6}{9} = {gray},
    cell{6}{11} = {gray},
    cell{7}{7} = {gray},
    cell{7}{8} = {gray},
    cell{7}{9} = {gray},
    cell{7}{10} = {gray},
    cell{7}{13} = {gray},
    cell{3}{10} = {lightgray},
    cell{4}{7} = {lightgray},
    cell{4}{8} = {lightgray},
    cell{4}{9} = {lightgray},
    cell{4}{10} = {lightgray},
    cell{6}{10} = {lightgray},
    cell{6}{12} = {lightgray},
    }   
        & & \SetCell[c=10]{c} $m$ & \\
        & & 20 & 25 & 30 & 35 & 40 & 50 & 75 & 100 & 150 & 200 & 250 \\
        \hline
        \SetCell[r=5]{l} $d$ & 
        9 & $\times$ & $\times$ & 0.6360 & 0.5264 & 0.4964 & $\times$ & $\times$ & 4.9010 &  &  &  \\
        & 12 & & $\times$ & $\times$ & 0.4414 & 2.0346 & 1.5257 & 1.8905 & 1.1872 &  & & \\
        & 15 & & & $\times$ & $\times$ & $\times$ & 0.3525 & 0.2949 & 0.2922 & 0.2883 & 0.2846  & \\
        & 20 & & & & $\times$ & $\times$ & $\times$ & $\times$ & 4.3245 & $\times$ & 3.5516 & \\
        & 30 & & & & & $\times$ & $\times$ & $\times$ & $\times$ & 0.1158 & 0.0923 & $\times$ \\
    \end{tblr}
    \caption{POD DEIM mean error. Failed simulations highlighted in dark gray, large mean errors highlighted in light gray.}
    \label{tab:POD_DEIM}
\end{table}

Tab.~\ref{tab:POD_DEIM} makes the robustness issues of the POD-DEIM apparent; these were also highlighted by~\cite{SolBraZab:2017:nsd,BraDavMer:2019:rmh}. For $19$ parameter combinations, at least one of the $500$ simulations fails to converge, and $7$ combinations yield unacceptable errors of $E_{\text{mean}}>1\%$. When the MOR scheme does converge to reasonable solutions, error levels are low, but this only happens for $11$ parameter combinations. Generally, higher values of $d$ and $m$ promote accuracy, but these trends are nonuniform. Additionally, there is no large region in the investigated algorithmic parameter space in which the POD-DEIM yields predictably robust results, such that it is not trivial to find parameters for which this method works as desired on the RVE problem considered here.

\begin{table}[h]
    \centering
    \begin{tblr}{
    colspec={cc|cccccccccc},
    cell{3}{3} = {gray},
    cell{3}{8} = {gray},
    cell{4}{4} = {gray},
    cell{4}{6} = {gray},
    cell{5}{5} = {gray},
    cell{5}{6} = {gray},
    cell{5}{7} = {gray},
    cell{5}{8} = {gray},
    cell{6}{6} = {gray},
    cell{6}{7} = {gray},
    cell{6}{8} = {gray},
    cell{7}{7} = {gray},
    cell{7}{8} = {gray},
    cell{7}{9} = {gray},
    cell{7}{13} = {gray},
    cell{3}{9} = {lightgray},
    cell{3}{10} = {lightgray},
    cell{4}{7} = {lightgray},
    cell{4}{8} = {lightgray},
    cell{4}{9} = {lightgray},
    cell{6}{10} = {lightgray},
    cell{6}{11} = {lightgray},
    cell{6}{12} = {lightgray},
    }   
        & & \SetCell[c=10]{c} $m$ & \\
        & & 20 & 25 & 30 & 35 & 40 & 50 & 75 & 100 & 150 & 200 & 250 \\
        \hline
        \SetCell[r=5]{l} $d$ & 
        9 & $\times$ & 0.7228 & 0.5178 & 0.5018 & 0.4785 & $\times$ & 3.1818 & 1.9488 &  &  &  \\
        & 12 & & $\times$ & 0.4452 & $\times$ & 1.1310 & 1.0947 & 1.0243 & 0.8432 &  & & \\
        & 15 & & & $\times$ & $\times$ & $\times$ & $\times$ & 0.2887 & 0.2847 & 0.2820  & \\
        & 20 & & & & $\times$ & $\times$ & $\times$ & 0.7315 & 1.9190 & 2.9067 & 2.4180 & \\
        & 30 & & & & & $\times$ & $\times$ & $\times$ & 0.1075 & 0.0852 & 0.0824 & $\times$ \\
    \end{tblr}
    \caption{POD LEHM mean error. Failed simulations highlighted in dark gray, large mean errors highlighted in light gray.}
    \label{tab:POD_LEHM}
\end{table}

In combination with the POD, the LEHM modification to the DEIM outlined above does not yield significant improvements. $13$ parameter combinations yield converging simulations with reasonable results, the error is unacceptably high in $8$ cases, and in $15$ cases at least one simulation fails to converge. Error levels and trends are very similar, and the desirable regions of the algorithmic parameter space are no more contiguous.

Meanwhile, the LPOD did not yield any parameter combinations for which all $500$ simulations converged in combination with DEIM-like hyperreduction methods, suggesting significant robustness issues\footnote{When not paired with any hyperreduction scheme, no robustness issues appeared. We attempted to couple the LPOD to the DEIM and LEHM, either with multiple localised, and with one global, hyperreduction model, to no avail. That the LPOD works as desired with the LSPG (see below) suggests that the specific combination of LPOD and DEIM is responsible for this lack of convergence.}.

\begin{table}[h]
    \centering
    \begin{tblr}{
    colspec={cc|cccccccccc},
    cell{3}{3} = {gray},
    cell{3}{5} = {gray},
    cell{3}{6} = {gray},
    cell{3}{7} = {gray},
    cell{3}{8} = {gray},
    cell{3}{9} = {gray},
    cell{3}{10} = {gray},
    cell{4}{4} = {gray},
    cell{4}{9} = {gray},
    cell{4}{10} = {gray},
    cell{5}{8} = {gray},
    cell{6}{6} = {gray},
    cell{6}{7} = {gray},
    cell{6}{8} = {gray},
    cell{7}{7} = {gray},
    cell{7}{8} = {gray},
    cell{7}{9} = {gray},
    cell{7}{13} = {gray},
    cell{5}{11} = {lightgray},
    cell{5}{12} = {lightgray},
    cell{6}{12} = {lightgray},
    cell{7}{12} = {lightgray},
    }   
        & & \SetCell[c=10]{c} $m$ & \\
        & & 20 & 25 & 30 & 35 & 40 & 50 & 75 & 100 & 150 & 200 & 250 \\
        \hline
        \SetCell[r=5]{l} $d$ & 
        9 & $\times$ & 0.3858 &$\times$&$\times$&$\times$&$\times$&$\times$& $\times$ &  &  &  \\
        & 12 & &$\times$& 0.3257 & 0.2852 & 0.2602 & 0.2487 &$\times$&$\times$&  & & \\
        & 15 & & & 0.2287 & 0.2228 & 0.2204 &$\times$& 0.1834 & 0.1784 & 2.2028 & 1.5465 & \\
        & 20 & & & &$\times$&$\times$&$\times$& 0.8375 & 0.7812 & 0.7827 & 1.4868 & \\
        & 30 & & & & &$\times$&$\times$&$\times$& 0.0628 & 0.0584 & 2.8437 &$\times$\\
    \end{tblr}
    \caption{PM LEHM mean error. Failed simulations highlighted in dark gray, large mean errors highlighted in light gray.}
    \label{tab:PM}
\end{table}

The PM, meanwhile, yields lower mean errors when all simulations converge, and there are only $4$ parameter combinations leading to excessive error levels. However, $18$ parameter combinations lead to at least one of $500$ simulations not converging; in particular, there is only one value of $m$ for which a PM model with $d=9$ runs robustly.

\begin{table}[h]
    \centering
    \begin{tblr}{
    colspec={cc|cccccccccc},
    cell{4}{4} = {gray},
    cell{4}{5} = {gray},
    cell{4}{6} = {gray},
    cell{5}{5} = {gray},
    cell{6}{6} = {gray},
    cell{6}{7} = {gray},
    cell{7}{7} = {gray},
    cell{7}{8} = {gray},
    cell{3}{9} = {lightgray},
    cell{3}{10} = {lightgray},
    cell{4}{9} = {lightgray},
    cell{4}{10} = {lightgray},
    cell{5}{12} = {lightgray},
    cell{6}{12} = {lightgray},
    cell{7}{11} = {lightgray},
    cell{7}{12} = {lightgray},
    cell{7}{13} = {lightgray},
    }   
        & & \SetCell[c=10]{c} $m$ & \\
        & & 20 & 25 & 30 & 35 & 40 & 50 & 75 & 100 & 150 & 200 & 250 \\
        \hline
        \SetCell[r=5]{l} $d$ & 
        9 & 0.2519 & 0.2254 & 0.2229 & 0.2193 & 0.2135 & 0.2090 & 1.1803 & 1.4237 &  &  &  \\
        & 12 & & $\times$ & $\times$ & $\times$ & 0.1789 & 0.1635 & 2.6949 & 2.6363 & & & \\
        & 15 & & & $\times$ & 0.2285 & 0.1804 & 0.1454 & 0.1309 & 0.1309 & 0.5261 & 2.3642 & \\
        & 20 & & & & $\times$ & $\times$ & 0.1575 & 0.1111 & 0.1018 & 0.0989 & 2.4503 & \\
        & 30 & & & & & $\times$ & $\times$ & 0.0946 & 0.0599 & 4.3670 & 3.4263 & 3.4195 \\
    \end{tblr}
    \caption{LLE LEHM mean error. Failed simulations highlighted in dark gray, large mean errors highlighted in light gray.}
    \label{tab:LLE}
\end{table}

The results obtained by the LLE-LEHM are more encouraging. The mean error levels are lower than those obtained by all other methods -- around half those obtained via the POD, and ca. $20-30\%$ lower than those obtained by the PM at many parameter combinations. While $9$ parameter combinations yield a mean error of $E_{\text{mean}}>1\%$, only $8$ parameter combinations yield one out of $500$ diverging simulations or more. Finally, there is a contiguous range of parameters in which the LLE-LEHM performs as desired in terms of accuracy and robustness on this problem, which means that it might be more straightforward to reliably select parameters for similar problems.

\begin{figure}[h]
    \centering
    \includegraphics[width=0.5\linewidth]{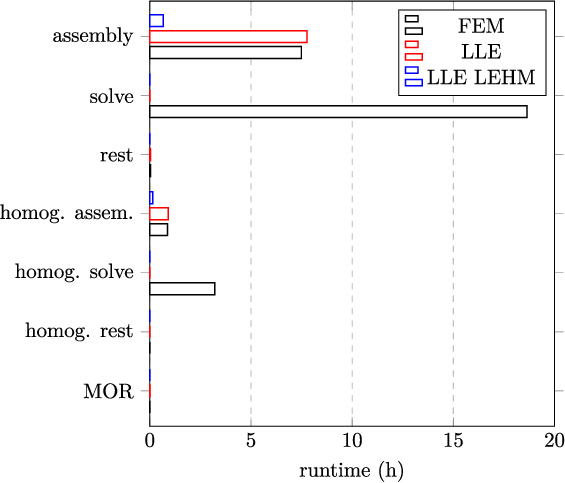}
    \caption{Runtime for LLE-LEHM with $d=15, m=100$.}
    \label{fig:runtime}
\end{figure}

Next, Fig.~\ref{fig:runtime} highlights the runtime consumed by different operations within the full-, reduced-, and hyperreduced FEM RVE homogenisation algorithms. Here, we only discuss results for the LLE-LEHM with $d=15$ and $m=100$, but similar observations hold for other methods and parameters.
As is to be expected, the runtime of the full FEM simulation is dominated by the linear solves in the Newton-Raphson scheme, followed closely, on account of the moderate problem size $D$, by the assembly. All other operations within the Newton-Raphson scheme only account for marginal computational costs. The runtime consumed by the homogenised stress- and stiffness computations break down similarly, though they account for less of the overall runtime since they only need to be performed once after the Newton-Raphson scheme converges. The Galerkin-reduced MOR scheme eliminates the cost of the linear solver nearly completely. Assembly costs increase slightly, since the stiffness matrix and residual still need to be assembled fully and since, for some macroscopic deformation gradient values $\ten{\bar{F}}$, one more iteration is required until convergence. The hyperreduced ROM slices these assembly costs significantly, though this reduction is not as severe as that in the cost of the linear solver operations. Thus, the assembly costs dominate the runtime of the hyperreduced model.

\begin{figure*}[h!]
\centering
\begin{subfigure}[t]{.45\textwidth}
  \centering
  \includegraphics[scale=0.9]{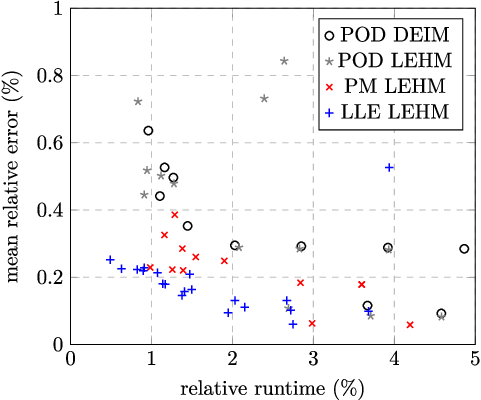}
  \caption{Mean relative error in displacement $\vec{u}$ over runtime relative to full FEM.}
  \label{fig:etu_DEIM}
\end{subfigure}%
\hspace{0.5cm}
\begin{subfigure}[t]{.45\textwidth}
  \centering
  \includegraphics[scale=0.9]{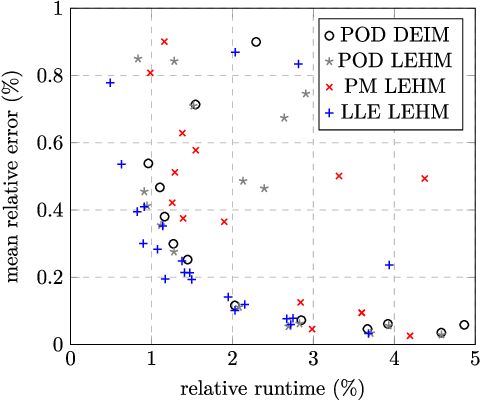}
  \caption{Mean relative error in homogenised stress $\ten{P}$ over runtime relative to full FEM}
  \label{fig:etp_DEIM}
\end{subfigure}
\caption{{Results with DEIM-like hyperreduction.} }
\label{fig:et_DEIM}
\end{figure*}

Next, Fig.~\ref{fig:et_DEIM} highlights the tradeoff between relative runtime reductions and relative errors (in the displacement $\vec{u}$ and the homogenised stress $\ten{\bar{P}}$) which can be achieved via different $d$ and $m$ for various MOR techniques coupled with DEIM-like hyperreduction methods. The LLE-LEHM Pareto-dominates the other techniques in the trade-off between runtime and accuracy particularly in $\vec{u}$. At a fixed computational budget, the LLE can roughly halve the error obtained by the POD and achieve an improvement of ca. $20\%$ over the PM, in addition to affecting robustness improvements. For example, a $50$-fold speedup can be achieved with a $0.3\%$ error using the POD, a $0.25\%$ error using the PM, and a $0.1\%$ error using the LLE. The advantage is palpable especially at low runtimes, and speedups of $200$ with mean errors of $E_{\text{mean}}<1\%$ (in particular, $E_{\text{mean}}\approx0.25\%$) can only be achieved using the LLE.

In the homogenised stress $\ten{\bar{P}}$, the advantage is less pronounced, as is to be expected since the averaging via Eq.~\eqref{eq:homogP} might mitigate local errors in the displacement. At low runtimes, there is a more significant reduction in the relative error; e.g., a $100$-fold speedup can be achieved with an error of around $0.3\%$ using the LLE and $0.5\%$ using the POD. The PM actually yields higher errors at a fixed computational budget than the POD, but this is mainly due to the PM not running robustly for smaller model sizes here.

Next, Tabs.~\ref{tab:POD_LSPG},~\ref{tab:LPOD_LSPG},~\ref{tab:PM_LSPG}, and~\ref{tab:LLE_LSPG} present the mean error $E_{\text{mean}}$ in the displacement $\vec{u}$ obtained by the POD, the LPOD, the PM and the LLE in tandem with the LSPG hyperreduction method. Again, parameter combinations for which at least one out of $500$ simulations failed to converge are highlighted in dark gray and parameter combinations yielding an excessively high mean error in light gray.

\begin{table}[h]
    \centering
    \begin{tblr}{
    colspec={cc|cccccccccc},
    cell{6}{6} = {lightgray},
    cell{6}{11} = {lightgray},
    cell{7}{13} = {lightgray},
    }   
        & & \SetCell[c=10]{c} $m$ & \\
        & & 20 & 25 & 30 & 35 & 40 & 50 & 75 & 100 & 150 & 200 & 250 \\
        \hline
        \SetCell[r=5]{l} $d$ & 
        9 & 0.7342 & 0.6632 & 0.6190 & 0.5944 & 0.5860 & 0.9419 & 0.7918 & 0.6996 &  &  &  \\
        & 12 & & 0.5502 & 0.4899 & 0.4731 & 0.6901 & 0.6401 & 0.5724 & 0.5320 &  & & \\
        & 15 & & & 0.4398 & 0.4425 & 0.3963 & 0.3965 & 0.3923 & 0.3881 & 0.3850 & 0.3786 & \\
        & 20 & & & & 1.3323 & 0.8790 & 0.6298 & 0.5070 & 0.7100 & 1.0957 & 0.8303 & \\
        & 30 & & & & & 0.2302 & 0.1913 & 0.1517 & 0.1429 & 0.1356 & 0.1323 & 1.0020 \\
    \end{tblr}
    \caption{POD LSPG mean error. Failed simulations highlighted in dark gray, large mean errors highlighted in light gray.}
    \label{tab:POD_LSPG}
\end{table}

The POD performs considerably more reliably in tandem with the LSPG than with DEIM-like hyperreduction methods on the considered example. While error levels are slightly higher for equivalent parameter combinations, no simulations fail to converge and only three parameter combinations yield mean error levels of $E_{\text{mean}}>1\%$. Performance trends are still not uniform in $d$ and $m$, but are far more predictable by comparison, making for more reliable parameter setting.

\begin{table}[h]
    \centering
    \begin{tblr}{
    colspec={cc|cccccccccc},
    cell{5}{5} = {lightgray},
    cell{5}{6} = {lightgray},
    cell{5}{7} = {lightgray},
    cell{5}{8} = {lightgray},
    cell{5}{9} = {lightgray},
    cell{5}{10} = {lightgray},
    cell{7}{10} = {lightgray},
    cell{7}{11} = {lightgray},
    cell{7}{12} = {lightgray},
    }   
        & & \SetCell[c=10]{c} $m$ & \\
        & & 20 & 25 & 30 & 35 & 40 & 50 & 75 & 100 & 150 & 200 & 250 \\
        \hline
        \SetCell[r=5]{l} $d$ & 
        9 & 0.3462 & 0.3374 & 0.3376 & 0.3307 & 0.3271 & 0.3258 & 0.3191 & 0.3141 &  &  &  \\
        & 12 & & 0.2854 & 0.2769 & 0.2748 & 0.2699 & 0.2672 & 0.5859 & 0.4999 &  & & \\
        & 15 & & & 2.4800 & 2.0267 & 1.6828 & 1.3331 & 1.2201 & 1.0216 & 0.7958 & 0.6923 &  \\
        & 20 & & & & 0.2315 & 0.2169 & 0.2040 & 0.1933 & 0.1928 & 0.6251 & 0.4960 & \\
        & 30 & & & & & 0.2230 & 0.1565 & 0.1263 & 1.4651 & 1.0175 & 1.0903 & 0.8995 \\
    \end{tblr}
    \caption{LPOD LSPG mean error. Failed simulations highlighted in dark gray, large mean errors highlighted in light gray.}
    \label{tab:LPOD_LSPG}
\end{table}

In tandem with the LSPG, the LPOD yields encouraging results. There are no diverging simulations for any parameter combination, and the mean error levels are significantly below those obtained via the POD, often by about $50\%$. $9$ parameter combinations yield mean errors of $E_{\text{mean}}>1\%$ and performance trends fluctuate more strongly in $d$ and $m$, meaning that parameter setting is not entirely straightforward.

\begin{table}[h]
    \centering
    \begin{tblr}{
    colspec={cc|cccccccccc},
    cell{3}{4} = {gray},
    cell{3}{5} = {gray},
    cell{3}{6} = {gray},
    cell{3}{7} = {gray},
    cell{3}{9} = {gray},
    cell{3}{10} = {gray},
    cell{4}{9} = {gray},
    }   
        & & \SetCell[c=10]{c} $m$ & \\
        & & 20 & 25 & 30 & 35 & 40 & 50 & 75 & 100 & 150 & 200 & 250 \\
        \hline
        \SetCell[r=5]{l} $d$ & 
        9 & 0.5979 & $\times$ & $\times$ & $\times$ & $\times$ & 0.7884 & $\times$ & $\times$ &  &  &  \\
        & 12 & & 0.3419 & 0.3410 & 0.3262 & 0.3114 & 0.3123 & $\times$ & 0.5107 &  & & \\
        & 15 & & & 0.2693 & 0.2603 & 0.2564 & 0.2559 & 0.2395 & 0.2346 & 0.4214 & 0.3744 & \\
        & 20 & & & & 0.2187 & 0.2216 & 0.2087 & 0.4496 & 0.3558 & 0.2858 & 0.5551 & \\
        & 30 & & & & & 0.1726 & 0.1279 & 0.1055 & 0.0980 & 0.0934 & 0.4669 & 0.6089 \\
    \end{tblr}
    \caption{PM LSPG mean error. Failed simulations highlighted in dark gray, large mean errors highlighted in light gray.}
    \label{tab:PM_LSPG}
\end{table}

The PM produces similar mean error levels as the LPOD; sometimes being outperformed and sometimes outperforming by some margin. The PM appears to have a higher accuracy ceiling in the investigated parameter range on the considered problem and performance trends seem slightly more predictably, but at least one out of $500$ simulations fails to converge for $7$ parameter combinations, mostly at $d=9$.

\begin{table}[h]
    \centering
    \begin{tblr}{
    colspec={cc|cccccccccc},
    }   
        & & \SetCell[c=10]{c} $m$ & \\
        & & 20 & 25 & 30 & 35 & 40 & 50 & 75 & 100 & 150 & 200 & 250 \\
        \hline
        \SetCell[r=5]{l} $d$ & 
        9 & 0.2543 & 0.2503 & 0.2504 & 0.2508 & 0.2504 & 0.2496 & 0.4951 & 0.5790 &  &  &  \\
        & 12 & & 0.2141 & 0.2086 & 0.2073 & 0.2072 & 0.4544 & 0.4452 & 0.3916 &  & & \\
        & 15 & & & 0.1884 & 0.1786 & 0.1726 & 0.1692 & 0.1645 & 0.1630 & 0.2552 & 0.5629 & \\
        & 20 & & & & 0.1670 & 0.1577 & 0.1501 & 0.1401 & 0.1365 & 0.1320 & 0.6165 & \\
        & 30 & & & & & 0.1469 & 0.1266 & 0.1007 & 0.0873 & 0.6937 & 0.5472 & 0.4516 \\
    \end{tblr}
    \caption{LLE LSPG mean error. Failed simulations highlighted in dark gray, large mean errors highlighted in light gray.}
    \label{tab:LLE_LSPG}
\end{table}

The performance of the LLE-LSPG is very encouraging: on the investigated problems, no simulations fail or yield a mean error level of $E_{\text{mean}}>1\%$. Additionally, the error in the displacement field $\vec{u}$ is lower than that obtained by other techniques, and performance trends are generally predictable.

\begin{figure*}[h]
\centering
\begin{subfigure}[t]{.45\textwidth}
  \centering
  \includegraphics[scale=0.9]{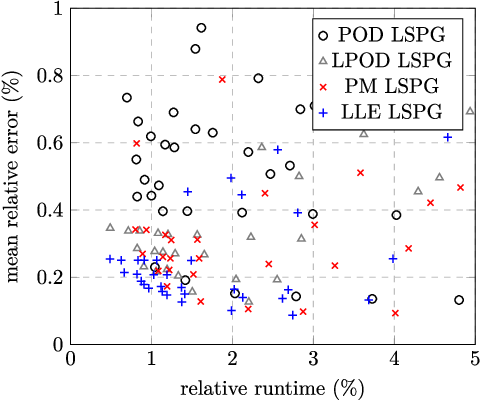}
  \caption{Mean relative error in displacement $\vec{u}$ over runtime relative to full FEM.}
  \label{fig:etu_LSPG}
\end{subfigure}%
\hspace{0.5cm}
\begin{subfigure}[t]{.45\textwidth}
  \centering
  \includegraphics[scale=0.9]{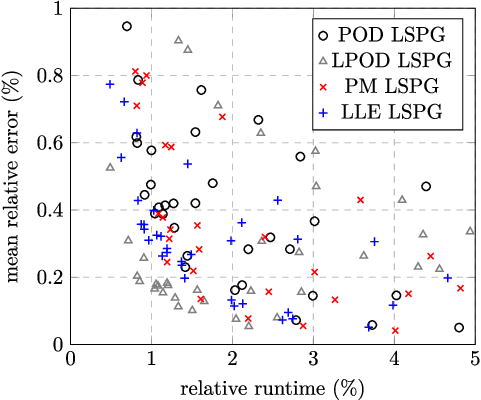}
  \caption{Mean relative error in homogenised stress $\ten{P}$ over runtime relative to full FEM}
  \label{fig:etp_LSPG}
\end{subfigure}
\caption{{Results with LSPG hyperreduction.} }
\label{fig:et_LSPG}
\end{figure*}

Finally, Fig.~\ref{fig:et_LSPG} highlights the tradeoff between mean errors in $\vec{u}$ and $\vec{\bar{P}}$ which can be achieved using different $d$ and $m$ using the POD, LPOD, PM, and LLE in combination with LSPG. Again, the LLE Pareto-dominates competing methods in the tradeoff between speed and accuracy in $\vec{u}$, outperforming the best of the alternatives by around $20-30\%$ in terms of relative error at a fixed computational budget. The LPOD and PM, meanwhile, perform very similarly on these measures. Again, the benefit conferred by LLE is particularly pronounced at low runtimes. A 200-fold speedup can be achieved with an error of around $0.25\%$, a 100-fold speedup with around $0.15\%$, and a 50-fold speedup with around $0.1\%$.

In terms of the error in $\ten{\bar{P}}$, the LLE slightly outperforms the POD and PM, but the LPOD Pareto-dominates here. The more continuously nonlinear approximation spaces obtained by the LLE and the PM thus do not seem to confer any advantage in the computation of the homogenised stress on the considered example. The observation made in the context of DEIM-like hyperreduction is accentuated here: continuously nonlinear approximation spaces yield an advantage when the underlying solution field is of interest, but this advantage is diminished when the homogenised quantities are more relevant.

\section{Summary and Outlook}\label{s:summary}

In a recent work~\cite{SchFau:2024:mla}, we proposed a projection-based nonlinear MOR scheme which uses graph-based manifold learning techniques to obtain flexible nonlinear approximation spaces. {In this work, we show how this approach can be employed to reduce computational costs by multiple orders of magnitude while retaining high levels of accuracy. To this end, } we extended the nonlinear MOR method with two hyperreduction methods: a DEIM-like approach and LSPG. Additionally, we improved the robustness and performance of the local online linearisation with an approximate Euler-backward-like scheme and a two-stage approach for the DEIM-like hyperreduction. Finally, the NLMOR scheme was extended to the homogenisation of stresses and stiffnesses based on RVE solutions. The resulting, hyperreduced algorithm no longer scales with the sizes $D$ and $|E|$ of the original problem.


On the example problem considered above, the hyperreduced manifold learning approach proposed in this work facilitates speedups of two orders of magnitude with negligible mean validation errors of $E_{\text{mean}}\approx0.1\%$ while requiring only $s=200$ snapshots.
When used in tandem with a DEIM-like hyperreduction scheme, the graph-based manifold learning approach Pareto-dominates alternative approaches in terms of the tradeoff between runtime and accuracy in the predicted displacement $\vec{u}$ and homogenised stress $\ten{\bar{P}}$. The advantage over competing methods is more pronounced in the displacement predictions, yielding a benefit of around $50\%$ over the POD and $20-30\%$ over the PM in the investigated RVE problem. The LPOD faces severe robustness issues, with at least one out of $500$ simulations diverging for every investigated algorithmic parameter combination. The POD and PM work as desired for a comparatively narrow and unpredictable parameter range, while the LLE yields good results for a somewhat broader, contiguous region in the algorithmic parameter space.

When used with LSPG, all methods perform much more robustly, though the mean error levels obtained with this method are slightly higher. The POD, LPOD, and PM yield unacceptably high mean errors or fail to converge only for $3$, $9$, and $7$ parameter combinations only, while the LLE never does. Again, the LLE Pareto-dominates all other methods in the tradeoff between speed and accuracy in the displacement $\vec{u}$, outperforming the best competing method by around $20-30\%$. In terms of the homogenised stress $\ten{\bar{P}}$, however, the LLE actually yields worse results than the LPOD.

For the example RVE considered here, nonlinear MOR techniques based on graph-based manifold learning methods such as LLE thus yield an advantage in the tradeoff between accuracy and speed especially when the underlying solution field is of interest, while the benefit to using such methods for the computation of the homogenised stress is less pronounced. 
It would be intriguing to investigate whether this observation generalises to other, more complex microstructures with larger phase contrasts and more nonlinear material behaviour. A systematic investigation over a range of RVEs could help identify where nonlinear approximation spaces might be employed profitably, and which methods to generate nonlinear approximation spaces might be most suitable for a particular class of multiscale problems.


Additionally, continuously and flexibly nonlinear approximation spaces might be of interest 
for history-dependent or multiphysical multiscale problems involving e.g. plasticity or damage indicators for which the evolution of variables on the microscale is critical. Of course, further developments are necessary to facilitate the application of these nonlinear MOR techniques to coupled and history-dependent problems, particularly if these involve localisation phenomena. A promising direction for research might be to generalise the manifold learning approaches explored in this work to multiple manifolds of coupled reduced solution variables. 

If these applications to more complicated problem classes prove promising, it would be sensible to attempt to push the nonlinear MOR techniques closer to their performance and development ceilings with further theoretical and implementational work, especially in terms of computational cost. With this in mind, it is important to note that the assembly, rather than linear system solutions, constitute the bottleneck in terms of runtime in the hyperreduced model.
Consequently, efforts toward further improving the speed of hyperreduced models ought to be targeted mainly at reducing $m$ rather than $d$. This to some extent motivates the considerable body of research effort expended on developing more advanced hyperreduction techniques. Note, however, that nonlinear Galerkin-MOR techniques, which work with smaller approximation spaces, might be able to yield improvements here, too. Smaller approximation spaces result in smaller reduced residual manifolds; and ideally fewer mesh entities $m$ being required to parameterise them.

\section*{Acknowledgements}

We would like to thank Rudy Geelen for his helpful comments.


\printbibliography

\end{document}